\documentclass[a4paper,twocolumn,10pt,accepted=2026-05-28]{quantumarticle}
\pdfoutput=1
\usepackage[utf8]{inputenc}
\usepackage[english]{babel}
\usepackage[T1]{fontenc}
\usepackage{amsmath}
\usepackage{hyperref}

\usepackage{tikz}
\usepackage{lipsum}

\usepackage{tikz}
\usepackage{qcircuit}

\usepackage{multirow}

\newcommand{\I}{\mathbbm{1}}

\newcommand{\tr}{\text{tr}}
\newcommand{\ket}[1]{| #1 \rangle}
\newcommand{\bra}[1]{\langle #1|}

\newcommand{\bracket}[3]{\langle #1|#2|#3 \rangle}

\newcommand{\be}{\begin{equation}}
\newcommand{\ee}{\end{equation}}
\newcommand{\bea}{\begin{eqnarray}}
\newcommand{\eea}{\end{eqnarray}}
\newcommand{\bes}{\begin{equation*}}
\newcommand{\ees}{\end{equation*}}
\newcommand{\beas}{\begin{eqnarray*}}
\newcommand{\eeas}{\end{eqnarray*}}

\newcommand{\x}{\mathrm{x}}
\newcommand{\ketbra}[1]{\ket{#1}\!\bra{#1}}
\newcommand{\dyad}[1]{\ket{#1}\!\bra{#1}}

\renewcommand{\H}{\mathcal{H}}

\def\x{\mathrm{x}}
\def\y{\mathrm{y}}

\def\F{\mathcal{F}}

\def\tr{\mathrm{tr}}

\DeclareMathOperator*{\Oplus}{\bigoplus}
\DeclareMathOperator*{\Otimes}{\bigotimes}

\begin{document}

\title{Detecting Entanglement by State Preparation and Local Measurements}

\author{Jaemin Kim}
\email{jeamink@es.aau.dk}
\affiliation{Department of Electronic Systems, Aalborg University, 9220 Aalborg, Denmark}
\affiliation{School of Electrical Engineering, Korea Advanced Institute of Science and Technology (KAIST), 291 Daehak-ro, Yuseong-gu, Daejeon 34141, 
Republic of Korea}
\orcid{0009-0007-1877-3107}

\author{Anindita Bera}
\email{aninditabera@bitmesra.ac.in}
\affiliation{Department of Mathematics, Birla Institute of Technology Mesra, Jharkhand 835215, India}

\author{Dariusz Chru\'sci\'nski}
\email{darch@fizyka.umk.pl}
\affiliation{Institute of Physics, Faculty of Physics, Astronomy, and Informatics, Nicolaus Copernicus University, Grudziadzka 5, 87-100 Torun, Poland}

\author{Joonwoo Bae}
\email{joonwoo.bae@kaist.ac.kr}
\affiliation{School of Electrical Engineering, Korea Advanced Institute of Science and Technology (KAIST), 291 Daehak-ro, Yuseong-gu, Daejeon 34141, Republic of Korea}

\maketitle

\begin{abstract}
Entanglement witnesses (EWs) are a collection of observables that can characterize separable states and, experimentally, estimating EWs can verify entangled states. In this work, we show that a fixed measurement setting on a multipartite entangled state, which we introduce as a \emph{network state} for the purpose, can estimate EWs. Namely, entangled states can be fully verified in a measurement-based manner, in which experimenters do not necessarily change measurement settings. We present a fixed measurement setting and network states for estimating decomposable EWs, equivalent to the partial transpose criteria. We also consider non-decomposable EWs that detect bound entangled states beyond the partial transpose criteria. The results can be extended to multipartite states such as graph states, a resource for measurement-based quantum computing, and readily applied to distributed settings such as quantum metrology or sensor networks where multipartite entangled states are resourceful.
\end{abstract}

\section{Introduction}

Entangled states are a key resource in quantum information processing, in particular, to achieve quantum advantages in computation and communication tasks. A higher level of security in cryptographic protocols \cite{PhysRevLett.98.230501}, higher level of quantum certification \cite{Pironio_2009, PhysRevLett.110.060405, PhysRevLett.108.200401, Bowles:2018aa} and higher network channel capacities \cite{PhysRevA.95.052329, PhysRevLett.125.150502} can be achieved by exploiting entangled states. Efficient computation with quantum resources can be realized by local measurements and entangled states only, called measurement-based quantum computation (MBQC) \cite{PhysRevLett.86.5188, Briegel:2009aa}.

Entanglement witnesses (EWs) \cite{TERHAL2000319, PhysRevA.62.052310} are versatile tools to detect entangled states both theoretically and experimentally. They correspond to observables that are non-negative for all separable states but negative for some entangled states \cite{HORODECKI19961, GUHNE20091, RevModPhys.81.865, Chru_ci_ski_2014, Friis:2019aa}. Hence, negative expectation values of EWs unambiguously conclude the presence of entangled states. In practice, multiple measurement settings may be necessary for estimating EWs.

In this work, we show that a fixed measurement setting can detect entangled states, together with the preparation of a multipartite state, which we call a {\it network state}. To be precise, we present a framework for detecting entangled states such that a fixed measurement over a network state $N$ and a state $\rho$ can find if the state $\rho$ is entangled, where a network state $N$ is obtained from an EW. Thus, entangled states can be verified in a measurement-based (MB) manner. In fact, the framework is closely connected to the activation of entanglement in that it finds if a state $\rho$ is entangled if a joint state $N\otimes \rho$ is more entangled than a state $N$~\cite{PhysRevLett.96.150501}.

We show how network states can be obtained from EWs. We present the construction of network states for decomposable EWs, which are equivalent to the partial transpose criteria, and also for non-decomposable EWs that detect bound entangled states beyond the partial transpose criteria, such as the Bell-diagonal EWs \cite{chruscinski2014class, PhysRevA.105.052401} from the Choi map \cite{Choi:1975aa} and its various generalizations \cite{PhysRevA.84.024302}, and the Breuer-Hall EW \cite{PhysRevLett.97.080501, Hall_2006}. The results are extended to multipartite systems: graph states \cite{PhysRevA.69.062311}, a resource for MBQC \cite{PhysRevLett.86.5188}, can be detected by network states and a fixed measurement.
{ 
We discuss the connection of our work to the entanglement activation \cite{PhysRevLett.96.150501} and the measurement-device-independent entanglement witnesses \cite{PhysRevLett.110.060405}.
}

{The importance of our results is threefold. Firstly, one can circumvent the need for precise control of measurement settings required to implement standard EWs. Note that imprecise control of measurement settings may cause errors in the realization of EWs. Our results show that preparing quantum states and performing fixed measurements are an experimentally feasible alternative to control of measurement settings, in a similar vein to MBQC which replaces quantum gates with state preparation and local measurements. Secondly, it is of fundamental interest to find that entangled states with a fixed measurement are a resource that can be replaced with observables. Note that MBQC shows that entangled states and local measurements can serve as an alternative for realizing quantum state transformations. Finally, the framework for estimating EWs in an MB manner finds various usefulness of distributed settings, such as distributed quantum metrology over networks, e.g., \cite{Chabuda:2020aa, Zhang_2021,Len:2022aa, PhysRevLett.121.043604}: estimation of joint observables enables distributed quantum information processing, e.g., \cite{PhysRevA.85.062326, Friis_2017, PhysRevLett.120.080501}.}
 
{
Experimental progress in preparing multipartite entangled resource states and implementing Bell-state-measurement-based protocols suggests that the present framework is relevant for near-term platforms. In particular, Smolin-type four-partite states have been realized experimentally~\cite{Amselem:2009aa, PhysRevLett.105.130501, PhysRevLett.109.040501}, and Bell-state measurements are a standard primitive in entanglement-swapping and networked quantum-information experiments~\cite{ma2012experimental}. Recently, entanglement pumping \cite{Dur_2007}, also a type of activation of entanglement, has been realized experimentally \cite{l43p-py55}. 
}

\section{Entanglement detection}
 
Let us begin by describing an experimental scenario of verifying entangled states, where the scenario is common in various experimental settings. We consider two quantum systems $A$ and $B$ on a Hilbert space $\H\otimes \H$. Interestingly, observables exist, denoted by $W=W^{\dagger}$, such that, for all separable states $\sigma_{\mathrm{sep}}$, 
\bea
\tr[W\sigma_{\mathrm{sep}}] \geq 0~~\mathrm{and }~~ \tr[W\rho_{\mathrm{ent}}] <0\nonumber
\eea
for some entangled states $\rho_{\mathrm{ent}}$ \cite{TERHAL2000319, PhysRevA.62.052310}. All entangled states can be verified by estimating some observables, which are called entanglement witnesses (EWs). In experiment, EWs can be estimated by altering measurement settings in general: negative expectation values unambiguously conclude entangled states \cite{GUHNE20091}. They can be also used to verify other properties, other than entanglement, such as a fidelity in the state preparation \cite{PhysRevA.76.030305, Haffner:2005aa}.






\begin{figure}[t]
\centering
\resizebox{1.0\linewidth}{!}{
\begin{tikzpicture}[scale=2]
    \definecolor{customBlue}{rgb}{0.2, 0.4, 1}
    \definecolor{customGreen}{rgb}{0.2, 1, 0.2}

\fill[lightgray, rounded corners=0.35cm] (-0.175,1.2) rectangle (0.175,-0.2);
\fill[lightgray, rounded corners=0.35cm] (-1-0.175,1.2) rectangle (-1+0.175,-0.2);
\fill[lightgray, rounded corners=0.35cm] (-0.5-0.7,-0.825) rectangle (-0.5+0.7,-1.175);
\draw[customGreen, ultra thick, dotted] (0,0) -- (0,-1);
\draw[customGreen, ultra thick, dotted] (-1,0) -- (-1,-1);

\draw[customGreen, ultra thick] (0.15,0) -- (-1+0.15,0);
\draw[customGreen, ultra thick] (0.15,-1) -- (-1+0.15,-1);

\draw[customBlue, ultra thick] (0.15,1) -- (-1+0.15,1);

    \fill[customBlue] (0,1) circle (0.15);
    \fill[customGreen] (0,0) circle (0.15);
    \fill[customGreen] (0,-1) circle (0.15);

    \fill[customBlue] (-1,1) circle (0.15);
    \fill[customGreen] (-1,0) circle (0.15);
    \fill[customGreen] (-1,-1) circle (0.15);

    \node at (-0.5,1.2) {$\rho^{(1)}$};
    \node at (-0.5,0.2) {$N^{(23)}$};

\node at (-1+0.25,0.75) {$A_1$};
\node at (-1+0.25,-0.25) {$A_2$};
\node at (-1+0.25,-1.25) {$A_3$};
\node at (+0.25,0.75) {$B_1$};
\node at (+0.25,-0.25) {$B_2$};
\node at (+0.25,-1.25) {$B_3$};

    \begin{scope}[xshift=1cm, yshift=-0.5cm, scale=0.7] 
        \fill[lightgray, rounded corners=0.25cm] (0,0) rectangle (0.35,1.4);

        \node at (0.6,0.7) {$=$};

        \node[draw, inner sep=3pt] at (2,0.7) {
            \Qcircuit @C=1em @R=1.2em {
            \lstick{} & \ctrl{1} & \gate{H} & \meter & \rstick{\hspace{-1.2em} 0} \\
            \lstick{} & \targ    & \qw      & \meter & \rstick{\hspace{-1.2em} 0} \\
            }
        };
    \end{scope}
\end{tikzpicture}
}
\caption{ A four-partite network state $N_{23}$ is prepared on $A_2A_3B_2B_3$ to detect an entangled state on $A_1B_1$. Bi-interactions and measurements in the computational basis can construct EWs, see also Eq. (\ref{eq:expew2}) in the text. Gray boxes denote a measurement in the basis $|\phi^{+}\rangle$, which is equivalent to a measurement in the computational basis after applying of a controlled-NOT gate followed by a Hadamard gate. }
\label{fig:fig1}
\end{figure}

\section{Measurement-Based Estimation}

{We here show that observables can be generally estimated in a measurement-based manner. We focus on EWs, a particular collection of observables for detecting entangled states. We devise multipartite quantum states, {\it called network states}, for estimating observables of multipartite quantum systems. } 

\subsection{Example }
\label{subsec:ex}
Let us begin with an illuminating example of an MB estimation of an EW. {Note that the proposed MB estimation applies to all observables, and we here focus on EWs for verifying entangled states.} Throughout, Bell states are denoted by $|\phi^{\pm}\rangle = (|00\rangle \pm |11\rangle)/\sqrt{2}$ and $|\psi^{\pm}\rangle = (|01\rangle \pm |10\rangle)/\sqrt{2}$. We consider an EW in the following,
\bea
W = \frac{1}{4} (\mathbbm{1} + X\otimes X + Y\otimes Y + Z\otimes Z) \label{eq:ewex}
\eea
where $X,Y$ and $Z$ are Pauli matrices. Eq. (\ref{eq:ewex}) shows that an EW can be estimated by various measurement settings. Note that a Bell state $|\psi^{-}\rangle$ is detected by an EW above.


The MB framework for estimating an observable in Eq. (\ref{eq:ewex}) can be achieved by a four-partite state, called a {\it network state}, constructed as follows, 
\bea
N_{23} &=& \frac{1}{4} |\psi^-\rangle_{A_2B_2} \langle \psi^-| \otimes |\phi^+\rangle_{A_3B_3} \langle \phi^+| + \nonumber \\
&&\frac{1}{12} (\mathbbm{1} - |\psi^-\rangle_{A_2B_2} \langle \psi^-|) \otimes (\mathbbm{1} - |\phi^+\rangle_{A_3B_3} \langle \phi^+|),~~~~ ~\label{eq:wr}
\eea
on sites $A_2A_3B_2B_3$, see Fig. \ref{fig:fig1}. {We then place a state $\rho$ of interest at $A_1B_1$, denoted by $\rho_{1}:=\rho^{( A_1B_1)}$, which we want to learn if it is entangled. Estimating an observable $W_1 :=W^{(A_1B_2)}$ with respect to the state $\rho_1$ can be equivalently seen in terms of the network state and a projection as follows},
\bea
&& \tr[\rho_1 W_1 ] =  \nonumber \\
&& 16~ \tr  [\rho_1\otimes N_{23} (\frac{1}{2}\mathbbm{1} - | \phi^+\rangle_{A_3B_3} \langle \phi^+|) \otimes P^{(12)}], \label{eq:ewr}
\eea
where $P^{(12 )} = |\phi^{+}\rangle_{A_1A_2}\langle \phi^+| \otimes |\phi^{+}\rangle_{B_1B_2}\langle \phi^+|$. {We write by the singlet fraction of a state $\tau$ as the overlap $\langle \phi^+| \tau |\phi^+\rangle$ with a maximally entangled state $|\phi^+\rangle$}, with which the relation above can be simplified as follows,
\bea
&& _{A_3 B_3}  \langle \phi^{+}| \tr_{12}[ \rho_1 \otimes N_{23} ~P^{(12)} ]  |\phi^{+}\rangle_{A_3 B_3} \nonumber \\ 
&&= \frac{1}{8} - \frac{1}{4}\tr[\rho_1 W_1].~~
\label{eq:expew2}
\eea
The first line in Eq. (\ref{eq:expew2}) can be obtained by preparing a network state followed by a fixed measurement. A state $\rho_1$ is entangled if the singlet fraction is greater than $1/8$ since $\tr[\sigma_{\mathrm{sep}}W ]\geq 0 $ in Eq. (\ref{eq:expew2}) for all separable states $\sigma_{\mathrm{sep}}$. Once a Bell measurement reports an outcome $P^{(12)} $, a singlet fraction is estimated from the probability of having outcome $|\phi^{+}\rangle$ on $A_3B_3$. Thus, it is shown that an EW can be estimated in a fully MB manner. 

{In fact, all EWs can be realized in an MB manner as shown above. Namely, one can construct a network state when an EW is provided, which is to be shown in Subsection \ref{sec:gc} in detail. The scheme to find a network state from an EW is also constructive.}


\begin{figure}[t]
    \centering
    \includegraphics[scale=.25 ]{ 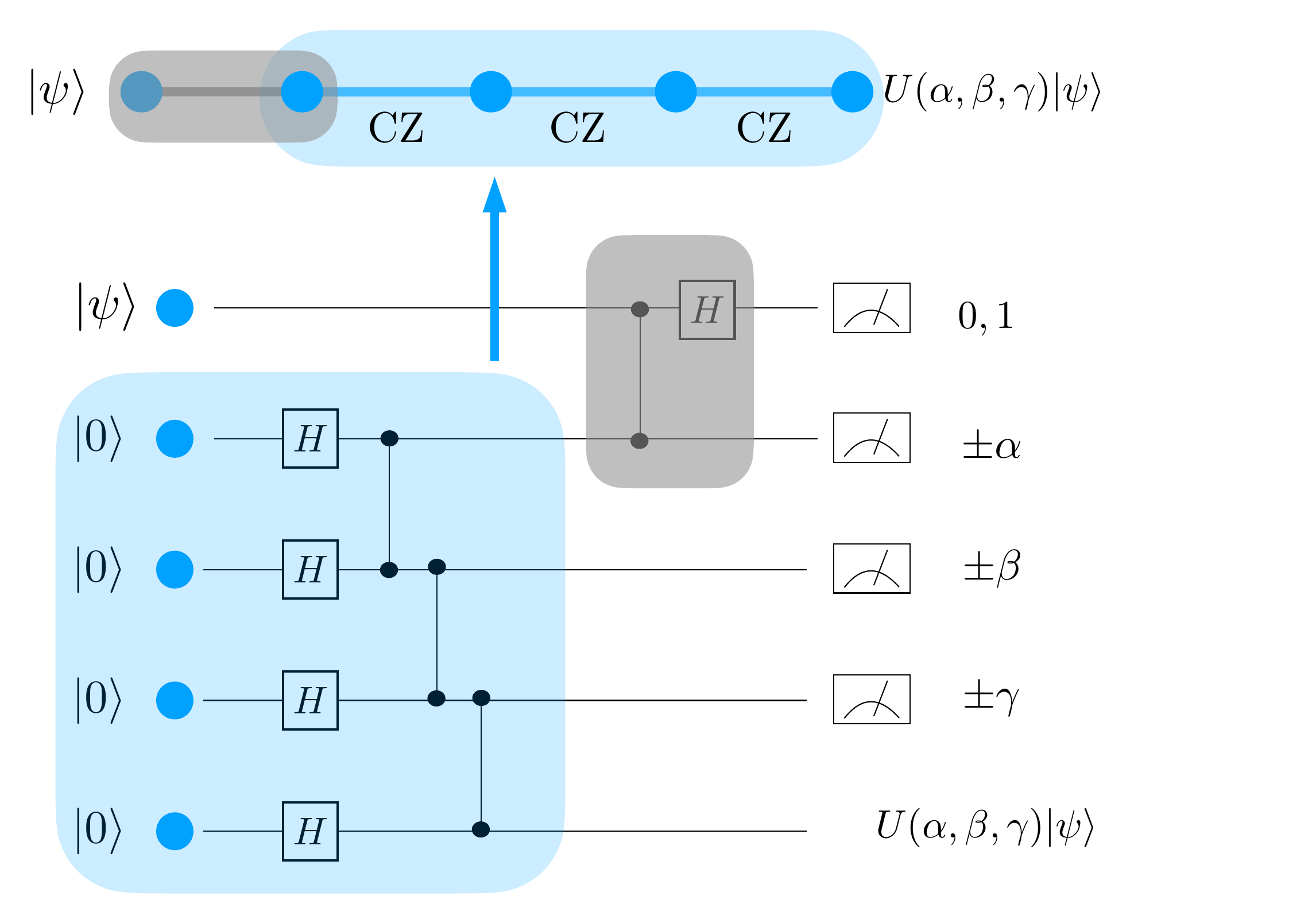}
    \caption{ { An elementary single-qubit gate operation in MBQC is realized by local measurements on a $5$-qubit cluster state, and the realization can be rephrased similarly to the MB framework for estimating EWs. (Blue box) Firstly, four qubits are prepared in a state $|+\rangle^{\otimes 4}$ by applying Hadamard gates  (H) on a state $|0\rangle^{\otimes 4}$. Applying CZ gates creates a $4$-qubit cluster state that is a resource. (Gray box) A CZ gate is applied between a state $|\psi\rangle$ and the $4$-qubit cluster state, preparing a $5$-qubit state. Then, local measurements are performed on the first four qubits in the basis $\{ (|0\rangle \pm e^{-i\theta}|1\rangle)/\sqrt{2}\}$ where $\theta=0, \alpha,\beta,\gamma$. An update of the measurement outcomes $(\alpha,\beta,\gamma)$ on the cluster state can implement a rotation gate on the last qubit. The resources needed to realize a single-qubit gate in MBQC can be summarized as $4$-qubit cluster states (blue box), CZ gates (gray box), and local measurements. The resources may be compared to the estimation scheme in Fig. \ref{fig:fig1}. 
    }}
    \label{fig:mbqc}
\end{figure}

\subsection{ Resources for the implementation} 
{
Experimental resources to realize the MB framework, i.e., the implementation of Eq. (\ref{eq:expew2}), are equivalent to those for MBQC. To clarify the equivalence, let us identify {\it resourceful states} that enable the MB tasks. Free resources are local measurements, that is, measurements on individual systems and classical communication that updates measurement outcomes to realize next quantum operations. }

Let us first consider the estimation of EWs in an MB manner, where resources are a network state and a joint measurement on bipartite systems, as shown in Fig. \ref{fig:fig1}. Note also that a network state is entangled, i.e., with respect to the bipartition $A_2A_3 : B_2B_3$. The detection scheme needs a measurement in the basis $|\phi^+\rangle$, which is equivalent to a controlled-NOT gate, a Hadamard gate, and a fixed-measurement in the $z$-direction; see also Fig. \ref{fig:fig1}. 

{Then, MBQC can be implemented by local measurements on a cluster state \cite{PhysRevLett.86.5188, Briegel:2009aa}. A cluster state, or generally a graph state, can be created by applying controlled-Z (CZ) gates on qubits including a system of interest, which one wants to transform, and ancillary qubits on which local measurements are to be performed. To this end, CZ gates are applied on a state $|\psi\rangle|+\rangle^{\otimes n}$ for some $n$ and a system state of interest $|\psi\rangle$. As it is shown Fig. \ref{fig:mbqc}, a cluster state for a single-qubit gate is generated by CZ gates on $|\psi\rangle |+\rangle^{\otimes 4}$. A single-qubit rotation on a state $|\psi\rangle$ can be implemented by local measurements. }

{ We here point out that resources of MBQC can be factorized into an $n$-qubit cluster state $|C_{n}\rangle$, created by CZ gates on $|+\rangle^{\otimes n}$, and an additional CZ gate, to be applied later to connect the cluster state to a state of interest $|\psi\rangle$. In Fig. \ref{fig:mbqc}, the former is in a blue box and the latter is in a gray box. We can identify $|C_n\rangle$ as a resource state and apply CZ gates on a state $|\psi\rangle |+\rangle^{\otimes n}$ to perform local measurements which realize MBQC. } 

{Therefore, we have shown that resources for MBQC and the framework of MB estimation are equivalent: large-size entangled states such as graph or network states, respectively, and additional controlled-$U$ gates connecting a resource state to a system state of interest. Similarly to graph states that have been experimentally realized \cite{PhysRevLett.95.210502, Lu:2007aa}, it is worth mentioning that a network state in Eq. (\ref{eq:wr}) is a variation of a Smolin state, a four-partite bound entangled state \cite{PhysRevA.63.032306}, which has been implemented in experiment \cite{ Amselem:2009aa, PhysRevLett.105.130501, PhysRevLett.109.040501}. Hence, a realization of the presented MB framework is readily available in an experiment. }

\subsection{ EWs via entanglement activation} \label{sec:act}

{We here observe that the construction of EWs in an MB manner can be equivalently seen as a form of activation of entanglement. If a state can activate the entanglement of another state, the state must be entangled, since separable states can activate none of the quantum states. In this subsection, we review the activation of entanglement in Ref. \cite{PhysRevLett.96.150501}.}

{ To be precise, one can define an entanglement measure as $E_d$, a maximal singlet fraction, for a state $N$ acting on a space $\H_{A_1A_2}\otimes \H_{B_1B_2}$, as follows
\bea
E_d[N] &=& \sup_{\F}~ \langle \phi_{d}^+| \F[ N ] |\phi_{d}^+\rangle ~\mathrm{for~a~local ~filtering~}\F   \nonumber \\
&& \F[N] = \frac{ K_{A}\otimes K_B N K_{A}^{\dagger}\otimes K_{B}^{\dagger} }{ \tr[  K_{A}^{\dagger}K_{A} \otimes K_{B}^{\dagger}K_{B} N ]},\label{eq:fil}
\eea
where $K_A: \H_{A_1A_2}\rightarrow \mathbbm{C}^d $ and $K_B: \H_{B_1B_2}\rightarrow \mathbbm{C}^d $. Then, a state $N$ having $E_d [N]\leq \eta$ may be activated by an entangled state $\sigma$ such that 
\bea
E_d[ \sigma\otimes N]>\eta, 
\label{eq:ac}
\eea
for some $\eta \in[1/d,1)$.} 

{All entangled states can be used to activate some other state: in Eq. (\ref{eq:ac}) a state $\sigma$ is entangled if and only if it can activate some other entangled state $N$ \cite{PhysRevLett.96.150501}. Hence, the phenomenon of activating entanglement can verify entangled states. From Eq. (\ref{eq:ac}), it is straightforward to derive a general form of an EW in terms of a network state $N := N^{(A_2 B_2 A_3 B_3)}$ and a parameter $\eta \in [1/d,1 )$, 
\bea
W_{A_2 B_2 } = \tr_{A_3B_3} \big[N (\eta \mathbbm{I} - |\phi_{d}^+ \rangle_{A_3B_3} \langle \phi_{d}^+ |)\big], \label{eq:ewm}
\eea
where $|\phi_{d}^+\rangle = \sum_{j=0}^{d-1}|jj\rangle /\sqrt{d}$. We reiterate that all EWs can be derived by the phenomenon of entanglement activation, and vice versa.}

\subsection{ Detection protocol from  entanglement activation}

{We exploit the scenario of activating entanglement as a framework for detecting entangled states and relate it to the verification of entangled states in an MB manner}. Namely, a state $\rho$ is entangled if it can be used to activate a state $N$. To this end, we consider a local filtering operation $\Phi$ as follows,
\begin{align}
\Phi (\rho \otimes N) = \tr_{A_1 B_1 A_2 B_2}[ P^{(12)} \rho^{(A_1 B_1)} \otimes N^{(A_2 B_2 A_3 B_3)}]. \nonumber
\end{align}
where $P^{(12)}$ is the projection onto maximally entangled states as follows,
\bea
P^{(12 )} = |\phi_{d}^{+}\rangle_{A_1A_2}\langle \phi_{d}^+| \otimes |\phi_{d}^{+}\rangle_{B_1B_2}\langle \phi_{d}^+|. \label{eq:projd} 
\eea
The singlet fraction after the filtering operation is given by,
\begin{align}
F_{\Phi}(\rho \otimes N) = \frac{ \bracket{\phi_d^+}{\Phi(\rho \otimes N)}{\phi_d^+} }{\tr[\Phi(\rho \otimes N)]}.\nonumber
\end{align}
With the specific local operation, the singlet fraction of a network state $N$ can be computed as, 
\begin{align}
\widetilde E_d(N) = \sup_{\sigma \in \mathrm{SEP}} F_{\Phi}(\sigma \otimes N),
\end{align}
where $\mathrm{SEP}$ denotes the set of separable states. It is clear that $\widetilde E_d(N) \le E_d(N)$ since a specific local operation is considered. Note also that for $\eta \in [\frac{1}{d},1)$, there exists a state $N$ such that $E_d(N) = \widetilde E_d(N) = \eta$ \cite{PhysRevLett.96.150501}. It holds that a state $\rho$ is entangled if we have $F_{\Phi}(\rho \otimes N) > \eta $ for $N$ such that $\widetilde E_d(N) \leq \eta$.

\subsection{ General construction of network states }
\label{sec:gc}

{We present a scheme for finding a network state when an EW is given. The scheme is constructive and applies to all EWs. It only relies on a decomposition of an EW in terms of positive operators. }

{Let us begin with a decomposition of an EW on $d\otimes d$ as follows,
\bea
W = \sum_{j} a_j W(j)^T ~\mathrm{with}~ W(j) \geq 0 ~\mathrm{and}~ a_j \in \mathbbm{R}. ~~~\label{eq:eeww}
\eea
Note that, clearly, a decomposition above is generally not unique. One can choose normalized non-negative operators $\{ \Pi(i) \geq 0\}$ and real constants $\{ c_j>0\}$ to construct a network state $N_{23}$,  
\bea
N_{23} &=&\sum_{j} c_j  W(j)_{A_2B_2} \otimes  \Pi(j)_{A_3B_3}.  \label{eq:ns} 
\eea
For instance, one can exploit general measurement operators, positive-operator-valued measure, for  $\{ \Pi(i) \geq 0\}$ up to a normalization.
}

{Note that coefficients $\{c_{j}\}$ in Eq. (\ref{eq:ns}) are determined by using the relation in Eq. (\ref{eq:ewm}). To be precise, we relate a network state $N$ in Eq. (\ref{eq:ns}) and an EW as follows, 
\bea
W_2 & =& ~ k \, \tr_{3} [N_{23}^{T_2}  ( \eta \mathbbm{1} - |\phi_{d}^+ \rangle_{A_3B_3} \langle \phi_{d}^+ |  )  ], \label{eq:ewn}
\eea
for some $\eta\geq 1/d$ and $k>0$; note here that a precise value $k$ is not related to the capability of detecting entangled states since it matters to find if the expectation value of $W_2$ is negative. From relations in Eqs. (\ref{eq:eeww}) and (\ref{eq:ns}), one can find $\{a_j\}$ and $\{c_j \}$ are related such that, $a_j = k \, c_j (\eta - \bracket{\phi_{d}^+}{\Pi(j)}{\phi_{d}^+})$.}

{Let us consider a state detected by an EW $W$. The detection scheme can be demonstrated by a network state, denoted by $\rho_1 = \rho^{(A_1B_1)}$ on sites $A_1B_1$. We have that
\bea
0&> & \tr[\rho W] = d^2  \tr[\rho_1 \otimes W_{2}^T P^{(12)}  ]  \nonumber\\
&\propto& \tr[\rho_1 \otimes N_{23} ~ P^{(12)} \otimes ( \eta \mathbbm{1} - |\phi_{d}^+ \rangle_{A_3B_3} \langle \phi_d^+ |  ) ]. ~~~~~ \label{eq:m}
\eea
From above, it follows that 
\bea
\eta &<&  _{A_3B_3}\langle \phi_{d}^+| \Lambda^{(1\rightarrow 3)} [\rho_1]|\phi_{d}^+\rangle_{A_3B_3} \nonumber \\
&&\mathrm{where}~ \Lambda^{( 1\rightarrow 3)} [\rho_1]= \frac{ \tr_{12}[\rho_{1}\otimes N_{23} P^{(12)}] }{ \tr[\rho_1\otimes N_{23} P^{(12)}]  }. ~~\label{eq:v}
\eea
Note that the construction above applies to all EWs in general. In Appendix~\ref{app1}, we reproduce a network state in Eq. (\ref{eq:wr}) by applying the general construction above.} \\

{{\it  Existence of network states for all EWs.} 
A straightforward way to construct a network state from an EW is to exploit positive and negative projections in its decomposition in Eq. (\ref{eq:ns}). Since an EW is not positive in general, it always has such a decomposition, from which one can find the existence of network states for all EWs.} For an EW, we find a decomposition,
\bea
 W = a_+ W_+^T - a_- W_-^T\nonumber
 \eea
with coefficients $a_{\pm} > 0$ and quantum states $W_{\pm}$ (i.e., $W_{\pm} \ge 0$ and $\tr[W_{\pm}]=1$). We choose $\eta = 1/d$ and find a network state as follows,
\begin{align}
N_{23} &=c_+ W_+ \otimes \frac{\I-\dyad{\phi^+_d}}{d^2-1} +  c_- W_- \otimes \dyad{\phi^+_d}.\nonumber
\end{align}
where
\bea
c_+ = \frac{(d-1)a_+}{(d-1)a_+ + a_-} ~~\mathrm{and} ~~c_-= \frac{a_-}{(d-1)a_+ + a_-}. \nonumber
\eea 
We have presented a straightforward one by exploiting a spectral decomposition of an EW. 
{ Note that an EW generally admits a decomposition above and, consequently, a network state for an EW can also be constructed accordingly.}
We also stress that a network state is not unique for an EW. 
\\

\begin{figure}[t]
    \centering
    \includegraphics[scale=.135 ]{ 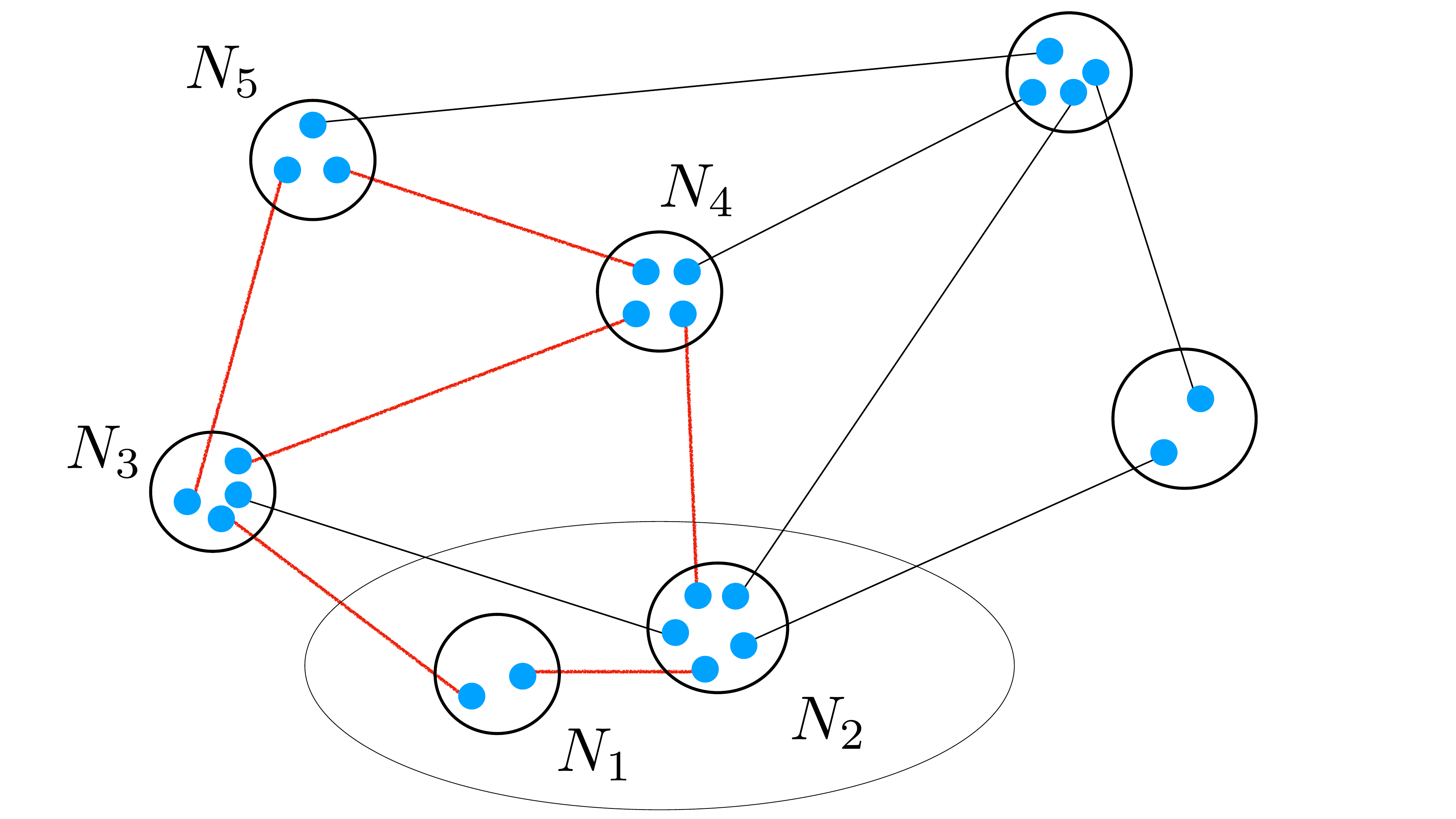}
    \caption{ { The framework of MB estimation can be extended to a multi-node network. A tripartite state $N_3N_5N_4$ can be manipulated to a bipartite state of $N_3N_4$. The framework of MB estimation applies to verify entanglement between nodes $N_3N_4$, for which a network state at nodes $N_1N_2N_3N_4$ can be exploited such that parties $N_1N_2$ estimate the singlet fraction and can find if the state $N_3 N_4$ is entangled. 
    }}
    \label{fig:network}
\end{figure}

\subsection{ { Applications in a quantum network }} \label{sec:tele}



{ The framework for estimating EWs with network states can be applied naturally in a quantum network that contains multiple nodes sharing large-sized entangled states.  In a network, one can identify a multipartite state of interest and also a network state, and detect entanglement of the state of interest remotely through the network state. For instance, as shown Fig. \ref{fig:network}, a state of interest may lie at nodes $N_3N_4$ resulting after a local measurement at $N_5$. It may be teleported to nodes $N_1N_2$ through a network state $N_3N_4N_1N_2$ and measurements at $N_3N_4$, where $N_1$ and $N_2$ may be closer and can estimate a singlet fidelity. Hence, entanglement shared between nodes $N_3N_4$ may be verified remotely at $N_1N_2$ by LOCC. }


{ The network protocol for detecting entangled states does not require varying measurement settings, which may be essential to a standard scenario of EWs. Fixed settings for Bell measurements and LOCC suffice to transfer the cost of varying measurement settings to the preparation of a network state, which may be in fact favorable or well-suited once quantum repeaters or entangled networks. }



{

\subsection{ Measurement-device-independent activation of entanglement} 
\label{semi MDI}

The proposed framework for detecting entangled states shares structural similarities with measurement-device-independent entanglement witnesses (MDI-EWs) \cite{PhysRevLett.108.200401,PhysRevLett.110.060405}. We discuss the framework from the MDI perspective and show how to relax assumptions about the measurements to realize the protocol in an MDI manner. 

The proposed framework contains measurements in the activation stage and the estimation of a singlet fraction. The assumption of Bell measurements may be relaxed, and the protocol can be realized in an MDI manner as follows. Let us consider a separable state $\sigma = \sum_k p_k \tau_k \otimes \xi_k$ and a threshold  $\eta$. Let $Q^{(A_1 A_2)}$ and $R^{(B_1B_2)}$ denote unknown positive-operator-valued-measure elements that may describe detection events of untrusted measurement devices. It suffices to show that $\sigma$ cannot increase the singlet fraction of $N$ larger than the threshold $\eta$. This means that, 
\bea
\tr[(Q^{(A_1 A_2)} \otimes R^{(B_1 B_2)} \sigma_1 \otimes W_{2}^T]  \geq 0 \nonumber
\eea
for all $Q$ and $R$, see also $W_{2}^T$ in Eq. (\ref{eq:ewn}). We then have: 
\bea
&& \sum_k p_k \tr[(Q^{(A_1 A_2)} \otimes R^{(B_1 B_2)}) (\tau_k^{(A_1)} \otimes \xi_k^{(B_1)}) W_{2}^T ] \nonumber \\
&\propto& \sum_k p_k \tr[(\widetilde{Q}^T \otimes \widetilde{R}^T) W] \ge 0 \nonumber 
\eea
since 
\begin{align*}
\widetilde{Q}^{(A_2)} &= \tr_{A_1}[Q^{(A_1 A_2)} (\sigma_k^{(A_1)} \otimes \I^{(A_2)})] \ge 0, \\
\widetilde{R}^{(B_2)} &= \tr_{B_1}[R^{(B_1 B_2)} (\xi_k^{(B_1)} \otimes \I^{(B_2)})] \ge 0\nonumber
\end{align*}
and an EW is non-negative for all separable states. Hence, we have shown that untrusted measurements on $A_1A_2$ and $B_1B_2$ can also unambiguously conclude entangled states, while the estimation of a singlet fraction is with trusted measurements. Hence, the protocol in the present form can be rephrased to be semi-MDI. 
}

{ In an MDI scenario, practical implementations of trusted quantum states may suffer from errors arising during state preparation. On the one hand, preparing a network state enables circumventing multiple error-prone steps to prepare a collection of trusted states in MDI-EWs. On the other hand, a single-copy network state is a multipartite entangled state, which may be more demanding to implement.
}
 
\section{Examples}

Let us apply the general construction of network states and present network states for EWs known so far, such as decomposable and non-decomposable instances, the latter of which are highly non-trivial in {quantum information theory}. Identifying all EWs is equivalent to characterizing the set of separable states, which is a challenging mathematical problem \cite{Stormer:1963aa}; the computational complexity also belongs to NP-Hard \cite{10.1145/780542.780545}. The construction of network states applies to all EWs. In what follows, we apply the general construction of network states to non-trivial EWs.

To this end, let $P_{st}$ denote a projection onto a Bell state, for $s,t = 0,\ldots, d-1$, in a dimension $d$,
\begin{align}
P_{st} = |\phi_{st}\rangle \langle \phi_{st}| ~\mathrm{where}~|\phi_{st} \rangle = \frac{1}{\sqrt{d}} \sum_{j = 0}^{d-1} \omega^{t j } \ket{j } \ket{j + s}, ~~ \label{eq:1}    
\end{align}
where $\omega = e^{2\pi i/d}$.
Projectors onto symmetric and anti-symmetric subspaces are denoted by $S_d$ and $A_d$, respectively,
\bea
S_d = \frac{\mathbbm{1+F} }{2}~~\mathrm{and} ~~A_d = \frac{\mathbbm{1 - F} }{2},
\eea
where $\mathbbm{F} = d P_{00}^{\Gamma}$ is a flip operator and $\Gamma$ denotes the partial transpose \cite{PhysRevA.61.062313}. Note that $\tr[A_d]=d(d-1)/2$ and $\tr[S_d] = d(d+1)/2$. Interestingly, high-dimensional Bell states and projections onto symmetric and anti-symmetric subspaces suffice to construct network states for known non-decomposable maps.

\subsection{ Decomposable EWs: the partial transposition }

Firstly, we consider a decomposable EW $W = Q^\Gamma$ for $Q\geq 0$ and $\tr[Q]=1$, for which a network state can be constructed as follows. We write by $\lambda:=\max_{i}|\lambda_i|$ where $\{\lambda_i\} $ are eigenvalues of an EW $W$ and a network state is obtained as,
\bea
N_{23}&=& c_1\left( \frac{\lambda\mathbbm{1} - Q^\Gamma }{\lambda d^2 - 1 }\right)^{(2)} \otimes P_{00}^{(3)} \nonumber\\
&&+ c_2 \left( \frac{ \lambda \mathbbm{1} +  Q^{\Gamma} }{\lambda d^2 + 1}\right)^{(2)} \otimes \left( \frac{\mathbbm{1} - P_{00}}{d^2 - 1}\right)^{(3)}, \label{eq:nsd}
\eea
where superscript $(j)$ stands for systems $A_jB_j$ and
\bea
c_1 = \frac{d^2 \lambda - 1}{d^3 \lambda  + d -2} ~~\mathrm{and}~ ~ c_2 =\frac{(d-1)(d^2 \lambda +1)}{d^3 \lambda  +d-2}. \nonumber
\eea
In the other way around, from a network state $N_{23}$ in Eq. (\ref{eq:nsd}) one can reproduce a decomposable EW, see Eq. (\ref{eq:ewm})
\bea
\tr_3[ N_{23 } (\frac{1}{d} \mathbbm{1} - |\phi_{00}\rangle_{A_3B_3}\langle \phi_{00}|)] = \frac{2(d-1)}{d(d^3 \lambda + d -2)} Q^\Gamma \propto Q^\Gamma.\nonumber
\eea
Hence, the partial transpose criteria \cite{PhysRevLett.77.1413} can be generally realized by preparing a network state with a fixed measurement.

As an instance, a network state for the decomposable and optimal EW $W = P_{00}^{\Gamma}$ can be found as
\bea
&& \frac{1}{d+2} \left( \frac{A_d}{\tr A_d} \right)^{(2)} \otimes P_{00}^{(3)} \nonumber \\
&&+\  \frac{d+1}{d+2} \left( \frac{S_d}{\tr S_d} \right)^{(2)} \otimes \left( \frac{\mathbbm{1}-P_{00}}{d^2-1} \right)^{(3)}. \nonumber
\eea
The network state above is known as a symmetric state being $UUVV^*$-invariant, and has been used to activate entanglement distillation with an infinitesimal amount of bound entanglement \cite{PhysRevLett.88.247901}.
The detailed construction is provided in Appendix~\ref{app:decEW}.

\subsection{ Non-decomposable EWs }

Secondly, to construct network states for non-decomposable EWs, we introduce paired Bell-diagonal (PBD) states as follows,
\bea
N_{23}^{(\mathrm{PBD})}(\Vec{\lambda}) = \sum_{s=0}^{d-1} \lambda_s ~\frac{1}{d} \sum_{t=0}^{d-1}  P_{st}^{(2)} \otimes P_{st}^{(3)}, \label{eq:PBD}
\eea
where $\Vec{\lambda}=(\lambda_0, \ldots, \lambda_{d-1})$ and $\sum_{s=0}^{d-1} \lambda_s = 1$.

\subsubsection*{ Bell-diagonal EWs}

A network state in Eq. (\ref{eq:PBD}) can be used to estimate expectation values of Bell-diagonal EWs \cite{chruscinski2014class},
 \bea
W[\Vec{\lambda}] = \sum_{s=0}^{d-1} \lambda_s \Pi_s - P_{00},~~\mathrm{where}~~ \Pi_s = \sum_{t=0}^{d-1} P_{st}.
\label{eq:Wa}
\eea
Note that the Choi map and its generalizations are well-known instances. Then, PBD network states construct Bell-diagonal EWs as follows,
\bea
\frac{\lambda_0}{d} W_2^T [\Vec{\lambda}] &=& \tr_3[ N_{23}^{(\mathrm{PBD})}(\Vec{\lambda}) (\lambda_0 \mathbbm{1} - |\phi_{00}\rangle_{A_3B_3}\langle \phi_{00}|)]. \nonumber
\eea
Hence, it is shown that all entangled states characterized by Bell-diagonal EWs can be detected by a fixed measurement and state preparation.

\subsubsection*{Choi EWs}

Instances of Bell-diagonal EWs for $d=3$ contain the Choi map \cite{Choi:1975aa} and its generalizations \cite{PhysRevA.84.024302, doi:10.1142/S1230161213500066}, that detect PPT entangled states. As it is shown in Eq. (\ref{eq:v}), once a filtering operation with a PBD network state in Eq. (\ref{eq:PBD}) is successful, entangled states are concluded by finding a singlet fraction. For the Choi map, entangled states are detected if the singlet fraction is larger than $2/3$. The proof is provided in Appendix~\ref{app:BDEW}.

\begin{figure}[t]
\centering
\resizebox{0.9\linewidth}{!}{
\begin{tikzpicture}[scale=0.5]
\definecolor{customBlue}{rgb}{0.2, 0.4, 1}
\definecolor{customGreen}{rgb}{0.2, 1, 0.2}
\begin{scope}
\clip (-4.5,-1.5) rectangle (10.5,7.5);
\begin{scope}
    \foreach \x in {-6,-3,...,12} {
        \draw[thick, black] (\x,-3) -- (\x,9);
    }
    \foreach \y in {-3,0,...,9} {
        \draw[thick,black] (-6,\y) -- (12,\y);
    }
\end{scope}

    \fill[lightgray, rounded corners=0.15cm] (0-0.35,3+0-0.35) rectangle (0+0.35,6+0+0.35);
    \fill[lightgray, rounded corners=0.15cm] (3-0.35,3+0-0.35) rectangle (3+0.35,6+0+0.35);
    \fill[lightgray, rounded corners=0.15cm] (6-0.35,3+0-0.35) rectangle (6+0.35,6+0+0.35);

    \draw[customGreen, ultra thick] (0,0) -- (3,0) -- (6,0);  
    \draw[customGreen, ultra thick] (0,3) -- (3,3) -- (6,3);
    \draw[customBlue, ultra thick] (0,6) -- (3,6) -- (6,6);
    \draw[white, ultra thick] (0,0) -- (0,3);
    \draw[white, ultra thick] (3,0) -- (3,3);
    \draw[white, ultra thick] (6,0) -- (6,3);
    \draw[customGreen, ultra thick, dotted] (0,0) -- (0,3);
    \draw[customGreen, ultra thick, dotted] (3,0) -- (3,3);
    \draw[customGreen, ultra thick, dotted] (6,0) -- (6,3);

    \fill[black] (-3,0) circle (0.3);
    \fill[black] (-3,3) circle (0.3);
    \fill[black] (-3,6) circle (0.3);

    \fill[black] (9,0) circle (0.3);
    \fill[black] (9,3) circle (0.3);
    \fill[black] (9,6) circle (0.3);

    \fill[customGreen] (0,0) circle (0.3);
    \fill[customGreen] (3,0) circle (0.3);
    \fill[customGreen] (6,0) circle (0.3);

    \fill[customGreen] (0,3) circle (0.3);
    \fill[customGreen] (3,3) circle (0.3);
    \fill[customGreen] (6,3) circle (0.3);

    \fill[customBlue] (0,6) circle (0.3);
    \fill[customBlue] (3,6) circle (0.3);
    \fill[customBlue] (6,6) circle (0.3);

    \node[fill=white, inner sep=0pt] at (1.5,1.5) {$N_{23}$};
    \node at (1.5,6.6) {$\rho_1$};
\end{scope}
\end{tikzpicture}
}
\caption{ A tripartite graph state can be detected by preparing a network state, Bell measurements, and a fixed measurement. Entangled states of arrayed qubits can be detected by state preparation and a fixed measurement. }
\label{fig:chain}
\end{figure}

\subsubsection*{ EWs from the Breuer-Hall map}

The Breuer-Hall (BH) map shown in Refs. \cite{PhysRevLett.97.080501, Hall_2006} derives highly non-trivial non-decomposable EWs,
\bea
\Lambda_{\textrm{BH}}(\rho) = \frac{1}{d-2} ( \tr(\rho) \mathbbm{1} - \rho - U \rho^T U^\dagger) \label{eq:BH map}
\eea
where $U$ is a skew-symmetric unitary operator satisfying $UU^\dagger = \mathbbm{1}$ and $U^T = -U$.
Then the BH EW is obtained as follows,
\bea
W_{\mathrm{BH}} = \frac{1}{d-2} ( \frac{1}{d}\mathbbm{1} - P_{00} - \frac{1}{d}\mathbbm{F}'), \label{eq:BH EW}
\eea
where $\mathbbm{F}' \equiv (\mathbbm{1} \otimes U) \mathbbm{F} (\mathbbm{1} \otimes U^\dagger)$. Note that the BH EW is optimal~\cite{PhysRevLett.97.080501, Hall_2006}.

A network state for the BH EW is obtained as follows,
\bea
N_{23}^{(\textrm{BH})}  
& = & c_0 \frac{1}{d^2} \sum_{s=0}^{d-1} \sum_{t=0}^{d-1} P_{st}^{(2)}  \otimes P_{st}^{(3)} \nonumber \\
&& + c_1 \left( \frac{\mathbbm{1} + \mathbbm{F}' }{d^2+d} \right)^{(2)} \otimes P_{00}^{(3)}   \nonumber \\
&& + c_2 \left( \frac{\mathbbm{1} - \mathbbm{F}' }{d^2-d} \right)^{(2)} \otimes \left( \frac{\mathbbm{1} - P_{00}}{d^2-1} \right)^{(3)}, \label{eq:nbhh}
\eea
where
\bea
c_0 = \frac{2d^2-2d}{3d^2 -3d +2},~\mathrm{and}~ c_1= \frac{d+1}{3d^2-3d+2}, \nonumber
\eea
and $c_2 = 1-c_0-c_1$. One can find that, from Eq. (\ref{eq:ewn})
\bea
W_{\mathrm{BH}}^T ~~\propto~~ \tr_3[ N_{23}^{(\mathrm{BH})}~ (\frac{1}{d} \mathbbm{1} - |\phi_{00}\rangle_{A_3B_3}\langle \phi_{00}|)]. \label{eq:bh}
\eea
Once a filtering operation in Eq. (\ref{eq:v}) is successful, entangled states are detected if a singlet fraction of a resulting state on $A_3B_3$ is larger than $1/d$.
See Appendix~\ref{app:BH} for the derivation.

\subsubsection*{Multipartite bound entangled states as a network state}

We also observe that a PBD state for $d=2$ with $\vec{\lambda} = (1/2,1/2)$ corresponds to a Smolin state \cite{PhysRevA.63.032306},
\bea
\rho_S = \frac{1}{4}\sum_{s,t=0,1 } P_{st}^{(A_2B_2)} \otimes P_{st}^{(A_3B_3)}. \nonumber
\eea
The state is invariant under permutations of $A_2A_3B_2B_3$ and remains PPT in any bipartite splitting: it is called a four-partite unlockable and undistillable entangled state. A Smolin state can be used to activate distillation of entanglement.

A Smolin state can be generalized to higher dimensions, with $\vec{\lambda} = (1/d,\ldots,1/d)$,
\bea
N_{23} (\Vec{\lambda}) = \frac{1}{d^2} \sum_{s=0}^{d-1} \sum_{t=0}^{d-1}  P_{st}^{(A_2 B_2)} \otimes P_{st}^{(A_3B_3)}. \nonumber
\eea
However, a Smolin state in a higher dimension $d>2$ no longer remains PPT in the bipartite splitting $A_2A_3:B_2B_3$. The network state then realizes an EW,
\bea
W = \frac{1}{d}\sum_{s=0}^{d-1}  \Pi_s - P_{00} = \frac{1}{d} \mathbbm{1} - P_{00} \label{reductionEW}
\eea
which is decomposable. It is also an EW that is derived from a reduction map \cite{PhysRevA.59.4206}. Note that a Smolin state corresponds to a network state that realizes a reduction EW for $d=2$.

\begin{figure}
\centering
\resizebox{0.7\linewidth}{!}{
\begin{tikzpicture}[scale=0.5]
    \definecolor{customBlue}{rgb}{0.2, 0.4, 1}
    \definecolor{customGreen}{rgb}{0.2, 1, 0.2}

    \fill[lightgray, rounded corners=0.15cm] (2-0.35,3+1-0.35) rectangle (2+0.35,6+1+0.35);
    \fill[lightgray, rounded corners=0.15cm] (6-0.35,3+1-0.35) rectangle (6+0.35,6+1+0.35);
    \fill[lightgray, rounded corners=0.15cm] (10-0.35,3+1-0.35) rectangle (10+0.35,6+1+0.35);

    \draw[customGreen, ultra thick] (0,0+0) -- (4,0+0) -- (8,0+0);  
    \draw[customGreen, ultra thick] (2,0+1) -- (6,0+1) -- (10,0+1); 
    \draw[customGreen, ultra thick] (0,0+0) -- (2,0+1);  
    \draw[customGreen, ultra thick] (4,0+0) -- (6,0+1);
    \draw[customGreen, ultra thick] (8,0+0) -- (10,0+1);

    \draw[customGreen, ultra thick] (2,3+1) -- (6,3+1) -- (10,3+1);

    \fill[lightgray, rounded corners=0.15cm] (0-0.35,3+0-0.35) rectangle (0+0.35,6+0+0.35);
    \fill[lightgray, rounded corners=0.15cm] (4-0.35,3+0-0.35) rectangle (4+0.35,6+0+0.35);
    \fill[lightgray, rounded corners=0.15cm] (8-0.35,3+0-0.35) rectangle (8+0.35,6+0+0.35);

    \draw[customGreen, ultra thick] (0,3+0) -- (4,3+0) -- (8,3+0);
    \draw[customGreen, ultra thick] (0,3+0) -- (2,3+1);
    \draw[customGreen, ultra thick] (4,3+0) -- (6,3+1);
    \draw[customGreen, ultra thick] (8,3+0) -- (10,3+1);
    
    \draw[customBlue, ultra thick] (0,6+0) -- (4,6+0) -- (8,6+0);
    \draw[customBlue, ultra thick] (2,6+1) -- (6,6+1) -- (10,6+1);
    \draw[customBlue, ultra thick] (0,6+0) -- (2,6+1);
    \draw[customBlue, ultra thick] (4,6+0) -- (6,6+1);
    \draw[customBlue, ultra thick] (8,6+0) -- (10,6+1);

    \fill[customGreen] (0,0+0) circle (0.3);
    \fill[customGreen] (4,0+0) circle (0.3);
    \fill[customGreen] (8,0+0) circle (0.3);
    \fill[customGreen] (2,0+1) circle (0.3);
    \fill[customGreen] (6,0+1) circle (0.3);
    \fill[customGreen] (10,0+1) circle (0.3);

    \fill[customGreen] (0,3+0) circle (0.3);
    \fill[customGreen] (4,3+0) circle (0.3);
    \fill[customGreen] (8,3+0) circle (0.3);
    \fill[customGreen] (2,3+1) circle (0.3);
    \fill[customGreen] (6,3+1) circle (0.3);
    \fill[customGreen] (10,3+1) circle (0.3);

    \fill[customBlue] (0,6+0) circle (0.3);
    \fill[customBlue] (4,6+0) circle (0.3);
    \fill[customBlue] (8,6+0) circle (0.3);
    \fill[customBlue] (2,6+1) circle (0.3);
    \fill[customBlue] (6,6+1) circle (0.3);
    \fill[customBlue] (10,6+1) circle (0.3);

    \draw[customGreen, ultra thick, dotted] (0,0+0) -- (0,3+0);
    \draw[customGreen, ultra thick, dotted] (4,0+0) -- (4,3+0);
    \draw[customGreen, ultra thick, dotted] (8,0+0) -- (8,3+0);
    \draw[customGreen, ultra thick, dotted] (2,0+1) -- (2,3+1);
    \draw[customGreen, ultra thick, dotted] (6,0+1) -- (6,3+1);
    \draw[customGreen, ultra thick, dotted] (10,0+1) -- (10,3+1);

    \node at (10,0+0.25) [text=customGreen] {\huge \(\cdot\)\(\cdot\)\(\cdot\)};
    \node[fill=white, inner sep=0pt] at (10,3+0.25) [text=customGreen] {\huge \(\cdot\)\(\cdot\)\(\cdot\)};
    \node at (10,6+0.25) [text=customBlue] {\huge \(\cdot\)\(\cdot\)\(\cdot\)};

    \node[fill=white, inner sep=0pt] at (1,1.75) {$N_{23}$};
    \node at (0.65,6.85) {$\rho_1$};
\end{tikzpicture}
}
\caption{The estimation of entanglement witnesses for multipartite graph states.}
\label{graph state scheme}
\end{figure}

\begin{figure*}[t]
    \centering
    \includegraphics[scale=.18]{ 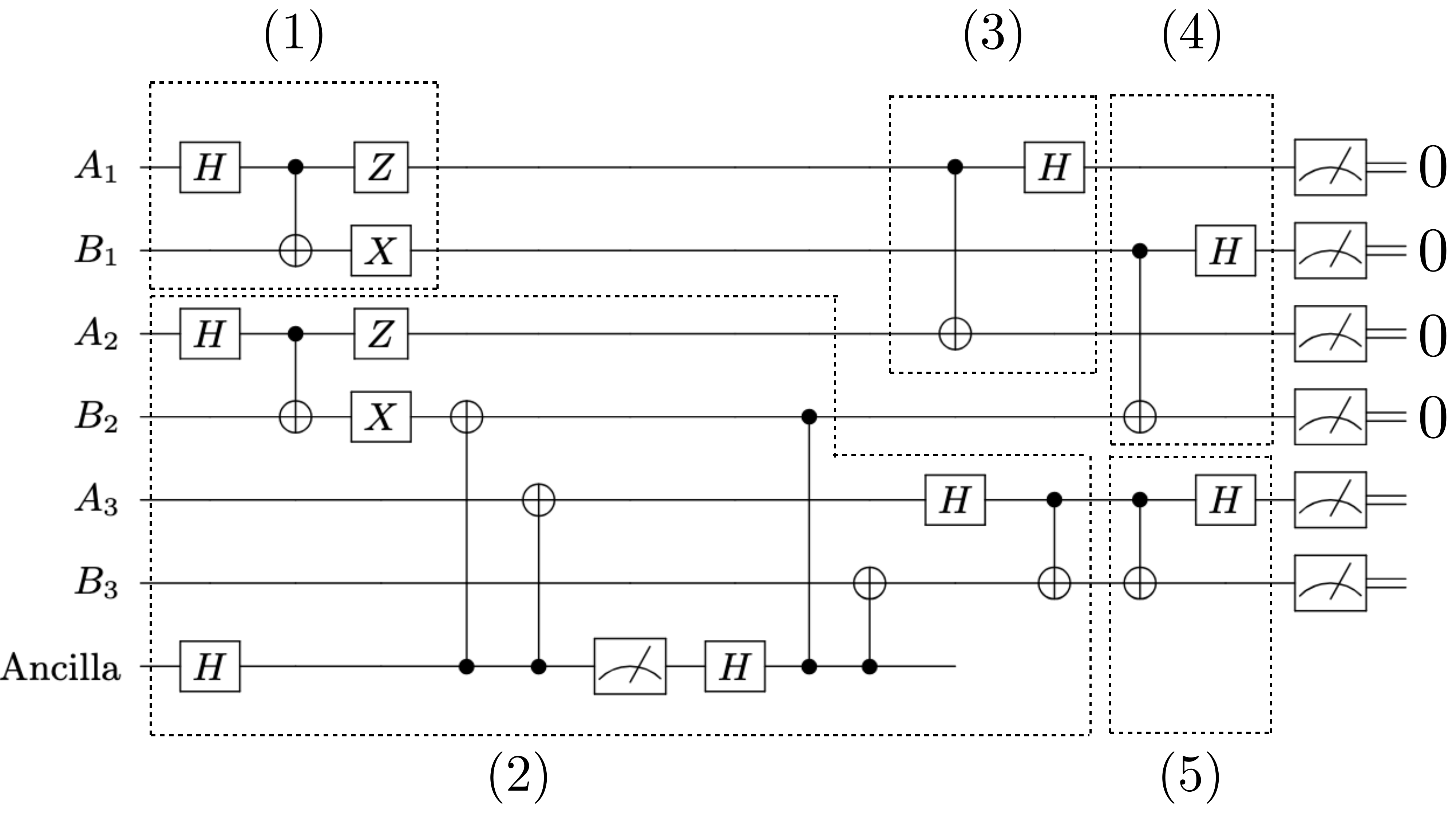}
    \caption{  {Estimation of an EW in Eq. (\ref{eq:ewex}) can be realized in a circuit, where a network state in Eq. (\ref{eq:wr}) is realized and measurements in the computational basis are applied, see also Eq. (\ref{eq:ewr}). (1) A state for detecting entanglement, $|\psi^{-}\rangle$, is generated in registers $A_1B_1$. (2) A network state from an EW  is prepared to detect entanglement generated in $A_1B_1$. (3) A projection onto a state $|\phi^+\rangle $ is realized by collecting outcomes $00$ in registers $A_1A_2$. (4) A projection onto a state $|\phi^+\rangle $ is realized by collecting outcomes $00$ in registers $B_1B_2$. (5) Given outcomes $0000$ in registers $A_1B_1A_2B_2$, the singlet fraction is estimated in $A_3B_3$: a state in register $A_1B_1$ is entangled if the probability of outcomes $00$ in $A_3B_3$ is larger than $1/2$.  }}
    \label{fig:4}
\end{figure*}

\subsection{ EWs for multipartite systems}

Thirdly, entanglement detection by state preparation can be extended to multipartite quantum states. We here, in particular, consider graph states, a class of states as a resource for MBQC \cite{PhysRevLett.86.5188}. Let us again present an instance for a three-qubit graph state, a Greenberger–Horne–Zeilinger (GHZ) state $|\psi\rangle = (|000\rangle +|111\rangle)/\sqrt{2}$ \cite{PhysRevA.62.062314}, see Fig. \ref{fig:chain}. An EW to detect a GHZ state may be given as,
\bea
W = \frac{1}{2} \mathbbm{1} - |\psi \rangle \langle \psi|.
\eea
A network state for an EW above can be constructed as,
\bea
N_{23} = \frac{1}{8} \sum_{a,b,c=0}^{1} \psi_{abc}^{(A_2 B_2 C_2)} \otimes \psi_{abc}^{(A_3 B_3 C_3)}.
\eea
where $\psi_{abc} = |\psi_{abc}\rangle \langle \psi_{abc}|$,
\bea
|\psi_{abc}\rangle = Z^a \otimes X^b \otimes X^c |\psi\rangle, ~~a,b,c \in \{0,1\}
\eea
with Pauli matrices $X$ and $Z$. It holds that
\bea
{}_{A_3B_3C_3} \langle \psi |\rho_1 \otimes N_{23} ~P^{(12)} |\psi \rangle_{A_3B_3C_3} = \frac{1}{16} - \frac{1}{8}\tr[\rho W], \nonumber
\eea
which shows detection of a genuinely multipartite entangled state $\rho$ by finding that the left-hand-side is greater than $1/16$. 

{The framework can be generalized to $n$-qubit graph states, see Fig. \ref{graph state scheme}. The detailed derivation is provided in Appendix~\ref{app:graph}.} \\

\subsection{ To construct non-decomposable EWs }

{Finally, let us investigate two entangled states defined by an EW, where one is an entangled state $\rho_1$ detected by an EW, and the other $N_{23}$ that realizes an EW for the detection in an MB manner, as shown in Eq. (\ref{eq:ewm}). Let us recall the result in Ref. \cite{PhysRevLett.96.150501}, see also Subsection \ref{sec:act}, that {\it an EW detects a set of entangled states which can activate its network state.} That is, an entangled state $\sigma$ activates $N$ such that $E_d [\sigma\otimes N]>\eta$ whereas $E_d[N] <\eta $, where $\eta \in [1/d,1)$. Then, we have $E_d [\sigma\otimes N]>1/d$ from which one finds that $\sigma\otimes N$ is distillable since $E_d$ computes the singlet fidelity \cite{PhysRevA.59.4206}, meaning that $\sigma\otimes N$ is not PPT, also signifying that either $\sigma$ or $N$ should be non-PPT. It is also clear that if both $\sigma$ and $N$ are PPT, then a state $\sigma\otimes N$ remains PPT under LOCC. Note that distillable entangled states must be non-PPT although the converse remains open; it has been a long-standing open question whether non-PPT states are distillable \cite{ PhysRevA.61.062313, PhysRevA.61.062312}. Stated equivalently, a PPT entangled state can activate a non-PPT state, but not another PPT state.} Hence, a network state $N_{23}$ to detect a PPT entangled state $\rho_1$ should be non-PPT. We thus conclude that multipartite non-PPT entangled states can construct non-decomposable EWs, which are highly non-trivial. Note that non-PPT states can construct both decomposable and nondecomposable EWs, as shown in the example of reduction EWs, see Eq. \eqref{reductionEW}.
\\

\section{Robustness}
\label{sec:robust}

The framework for detecting entangled states in the subsection \ref{sec:gc} contains elements, the preparation of a network state and a fixed measurement for estimating a singlet fraction. We reiterate the relation between the estimation of an EW and the framework with a network state, 
\begin{align}
\tr[\rho W] = \alpha ~ \tr[\rho_1 \otimes N_{23} P^{(12)} (\eta \I - |\phi_{d}^+\rangle \langle \phi_{d}^+ | )_3 ],
\end{align}
for some $\alpha>0$. We here investigate the effect of noise on the preparation of a network state and a fixed measurement and show that the framework can unambiguously detect entangled states. That is, for separable states, the framework with noisy resources gives a non-negative expectation value.

\subsubsection*{ Noisy network states }
Suppose that the network states $N$ is noisy such that
\begin{align*}
\widetilde N &= (1-p) N + p \frac{\I}{d^{2n}},
\end{align*}
with some noise fraction $p$. It follows that,
\bea
&&\alpha~ \tr[\rho_1 \otimes \widetilde N_{23} P^{(12)} (\eta \I -  |\phi_{d}^+\rangle \langle \phi_{d}^+ | )_3 ] \nonumber \\
&=& (1-p) \tr[\rho W] + \alpha p \frac{1}{d^{2n}} \frac{1}{d^{n}} (\eta d^n-1) \nonumber \\
&\geq & (1-p) \tr[\rho W]. \nonumber
\eea
Note that the term apart from $\tr[\rho W]$ in the second line is not negative since $\eta\geq 1/d$. Hence, noisy network states can be used to detect entangled states unambiguously.

\subsubsection*{White noise in singlet fraction estimation}
Suppose that a fixed measurement setting for estimating a singlet fraction is noisy,
\begin{align}
(1-q)  |\phi_{d}^+\rangle \langle \phi_{d}^+ | + q \frac{\I}{d^n} \nonumber
\end{align}
with some noise fraction $q$. It follows that
\bea
&& \alpha ~ \tr[\rho_1 \otimes N_{23} P^{(12)} (\eta \I - (1-q)  |\phi_{d}^+\rangle \langle \phi_{d}^+ | - q \frac{\I}{d^n}  )_3 ]  \nonumber \\
 &= & (1-q)\tr[\rho W] + \alpha q \frac{1}{d^n} (\eta - \frac{1}{d^n} \tr[\rho^T N] ) \nonumber \\
 &\geq & (1-q) \tr[\rho W]. \nonumber
\eea
Thus, when a fixed measurement setting is noisy, the framework gives non-negative expectation values for separable states.

\section{Realization in quantum circuits}

The framework of detecting entangled states by a fixed measurement can be realized in a quantum circuit. Here, we demonstrate the realization of a network state and the estimation of an EW with an example in Eq. (\ref{eq:ewex}). To facilitate the construction of a quantum circuit, we exploit a decomposition of the network state in the following, 
\begin{align}
N_{23} &= \frac{1}{4} \big( \ketbra{\psi^-}_{A_2 B_2} \otimes \ketbra{\phi^+}_{A_3 B_3}  \nonumber \\
&~~~~~~ + \ketbra{\psi^+}_{A_2 B_2} \otimes \ketbra{\psi^+}_{A_3 B_3} \nonumber \\
&~~~~~~ + \ketbra{\phi^-}_{A_2 B_2} \otimes \ketbra{\phi^-}_{A_3 B_3} \nonumber \\
&~~~~~~ + \ketbra{\phi^+}_{A_2 B_2} \otimes \ketbra{\psi-}_{A_3 B_3} \big), 
\end{align}
which is identical to the state in Eq. \eqref{eq:wr}. Then, a circuit for realizing the network state is shown in Fig. \ref{fig:4}, where blocks (1)-(5) describe realizations of a network state, projection on a state $|\phi^+\rangle$, measurements of estimating the singlet fraction. We remark that a fixed measurement setting is used for estimating an EW. 

\begin{figure}[t]
\centering
\includegraphics[width=\linewidth]{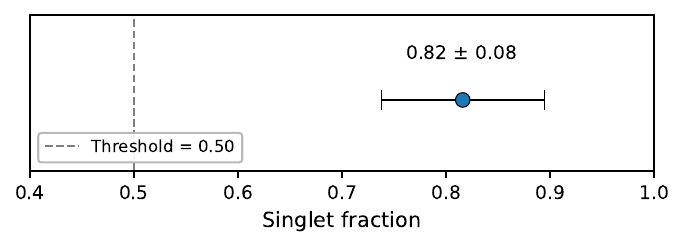}
\caption{
Proof-of-principle demonstration of the circuit in Fig.~\ref{fig:4} has been shown. 
The circuit was run on ibm\_marrakesh from 2026-03-23 to 2026-03-26. The singlet fraction is estimated from $10$ independent runs of $10,000$ shots each. Note that the threshold is $1/2$ with a dashed line. The average probability is $0.82$ and the standard deviation is obtained $0.08$. 
Hence, the results confirm the detection of entangled states on a noisy quantum processor.
}
\label{fig:result}
\end{figure}

Considering the effect of noise in Section \ref{sec:robust}, one may expect that the framework can be realized in a noisy implementation of quantum information processing. We then implemented the circuit in the { ibm\_marrakesh in March, 2026}. The results are presented in Fig. \ref{fig:result}, demonstrating the robustness of the proposed framework for estimating EWs in a realistic setting. Although noise is present, the estimation of an EW shows a singlet fraction {$0.82$} higher than the threshold $1/2$, detecting the presence of an entangled state in registers $A_1B_1$.

\section{Conclusion}

In conclusion, we have established an MB framework for estimating observables, in particular EWs, and shown the construction of network states for observables. It is also of fundamental interest to find that entanglement demonstrates elements of quantum theory, both dynamics, i.e., MBQC and observables. Our results shed new light on detecting and manipulating entangled states in a realistic scenario where precise controls via quantum gates are yet limited, such as an array of superconducting qubits, neutral atoms, or photons, where the preparation of a multipartite state and a fixed measurement are readily experimentally feasible. The results also open a new avenue of exploiting entangled states in distributed quantum information processing such as computing and metrology where entanglement is resourceful and quantum estimation {\it per se} corresponds to computing or metrology. 

In future directions, it would be interesting to realize the estimation of EWs via network states experimentally by considering the effects of noise in a realistic setting. Note that some network states, such as Smolin, have been realized in experiment \cite{Amselem:2009aa, PhysRevLett.105.130501, PhysRevLett.109.040501}. It is also interesting to exploit network states in a network and security scenario such as a measurement-device-independent scenario, see Refs. \cite{PhysRevLett.108.200401,PhysRevLett.110.060405}, so that observables such as EWs are estimated by relaxing assumptions on joint measurements; estimation of EWs, as well as detection of entanglement, is achieved in a higher level of robustness against imperfect implementations. 
{ In addition, it would be interesting to extend the framework with a network state to the verification of other quantum resources other than entanglement, such as quantum steering~\cite{Cavalcanti2013EntanglementSteering,zhao2020experimental} and quantum memories~\cite{Rosset2018QuantumMemories}.}

\section*{Acknowledgement}

J.K.\ and J.B.\ were supported by the National Research Foundation of Korea (RS-2025-00561467) and the Institute for Information \& Communication Technology Promotion (IITP) (RS-2023-00229524, RS-2025-02304540, RS-2025-25464876, RS-2025-25464616). J.K.\ was supported, in part, by the Danish National Research Foundation(DNRF), through the Center of Excellence CLASSIQUE, grant nr. DNRF187. DC was supported by the Polish National Science Center project No. 2024/55/B/ST2/01781. 
A.B.\ acknowledges the support received for this research from the research grant sanctioned by the National Board for Higher Mathematics (NBHM), Department of Atomic Energy (DAE), Government of India, with sanction letter no: 02011/32/2025/NBHM(R.P)/R\&D II/9677; 
the funding under the ARG program from Anusandhan National Research Foundation (ANRF) with file number: ANRF/ARG/2025/004617/PS;
under the seed money scheme from Birla Institute of Technology Mesra with sanction letter no: DRIE/SMS/DRIE-10917/2025-26/3857.

\section*{Data availability}
{ The experimental raw data from the IBM quantum device are publicly available at \url{https://doi.org/10.5281/zenodo.19337583}.}

\bibliographystyle{quantum}
\bibliography{bibedsp}

\begin{thebibliography}{10}

\bibitem{PhysRevLett.98.230501}
Antonio Ac\'{\i}n, Nicolas Brunner, Nicolas Gisin, Serge Massar, Stefano Pironio, and Valerio Scarani.
\newblock ``Device-independent security of quantum cryptography against collective attacks''.
\newblock \href{https://dx.doi.org/10.1103/PhysRevLett.98.230501}{Phys. Rev. Lett. {\bf 98}, 230501}~(2007).

\bibitem{Pironio_2009}
Stefano Pironio, Antonio Ac{\'\i}n, Nicolas Brunner, Nicolas Gisin, Serge Massar, and Valerio Scarani.
\newblock ``Device-independent quantum key distribution secure against collective attacks''.
\newblock \href{https://dx.doi.org/10.1088/1367-2630/11/4/045021}{New Journal of Physics {\bf 11}, 045021}~(2009).

\bibitem{PhysRevLett.110.060405}
Cyril Branciard, Denis Rosset, Yeong-Cherng Liang, and Nicolas Gisin.
\newblock ``Measurement-device-independent entanglement witnesses for all entangled quantum states''.
\newblock \href{https://dx.doi.org/10.1103/PhysRevLett.110.060405}{Phys. Rev. Lett. {\bf 110}, 060405}~(2013).

\bibitem{PhysRevLett.108.200401}
Francesco Buscemi.
\newblock ``All entangled quantum states are nonlocal''.
\newblock \href{https://dx.doi.org/10.1103/PhysRevLett.108.200401}{Phys. Rev. Lett. {\bf 108}, 200401}~(2012).

\bibitem{Bowles:2018aa}
Joseph Bowles, Ivan {\v S}upi{\'c}, Daniel Cavalcanti, and Antonio Ac{\'\i}n.
\newblock ``Device-independent entanglement certification of all entangled states''.
\newblock \href{https://dx.doi.org/10.1103/PhysRevLett.121.180503}{Physical Review Letters {\bf 121}, 180503--}~(2018).

\bibitem{PhysRevA.95.052329}
Yihui Quek and Peter~W. Shor.
\newblock ``Quantum and superquantum enhancements to two-sender, two-receiver channels''.
\newblock \href{https://dx.doi.org/10.1103/PhysRevA.95.052329}{Phys. Rev. A {\bf 95}, 052329}~(2017).

\bibitem{PhysRevLett.125.150502}
Jiyoung Yun, Ashutosh Rai, and Joonwoo Bae.
\newblock ``Nonlocal network coding in interference channels''.
\newblock \href{https://dx.doi.org/10.1103/PhysRevLett.125.150502}{Phys. Rev. Lett. {\bf 125}, 150502}~(2020).

\bibitem{PhysRevLett.86.5188}
Robert Raussendorf and Hans~J. Briegel.
\newblock ``A one-way quantum computer''.
\newblock \href{https://dx.doi.org/10.1103/PhysRevLett.86.5188}{Phys. Rev. Lett. {\bf 86}, 5188--5191}~(2001).

\bibitem{Briegel:2009aa}
H.~J. Briegel, D.~E. Browne, W.~D{\"u}r, R.~Raussendorf, and M.~Van~den Nest.
\newblock ``Measurement-based quantum computation''.
\newblock \href{https://dx.doi.org/10.1038/nphys1157}{Nature Physics {\bf 5}, 19--26}~(2009).

\bibitem{TERHAL2000319}
Barbara~M. Terhal.
\newblock ``Bell inequalities and the separability criterion''.
\newblock \href{https://dx.doi.org/10.1016/S0375-9601(00)00401-1}{Physics Letters A {\bf 271}, 319--326}~(2000).

\bibitem{PhysRevA.62.052310}
M.~Lewenstein, B.~Kraus, J.~I. Cirac, and P.~Horodecki.
\newblock ``Optimization of entanglement witnesses''.
\newblock \href{https://dx.doi.org/10.1103/PhysRevA.62.052310}{Phys. Rev. A {\bf 62}, 052310}~(2000).

\bibitem{HORODECKI19961}
Micha{\l} Horodecki, Pawe{\l} Horodecki, and Ryszard Horodecki.
\newblock ``Separability of mixed states: necessary and sufficient conditions''.
\newblock \href{https://dx.doi.org/10.1016/S0375-9601(96)00706-2}{Physics Letters A {\bf 223}, 1--8}~(1996).

\bibitem{GUHNE20091}
Otfried G{\"u}hne and G{\'e}za T{\'o}th.
\newblock ``Entanglement detection''.
\newblock \href{https://dx.doi.org/10.1016/j.physrep.2009.02.004}{Physics Reports {\bf 474}, 1--75}~(2009).

\bibitem{RevModPhys.81.865}
Ryszard Horodecki, Pawe\l{} Horodecki, Micha\l{} Horodecki, and Karol Horodecki.
\newblock ``Quantum entanglement''.
\newblock \href{https://dx.doi.org/10.1103/RevModPhys.81.865}{Rev. Mod. Phys. {\bf 81}, 865--942}~(2009).

\bibitem{Chru_ci_ski_2014}
Dariusz Chru{\'{s}}ci{\'{n}}ski and Gniewomir Sarbicki.
\newblock ``Entanglement witnesses: construction, analysis and classification''.
\newblock \href{https://dx.doi.org/10.1088/1751-8113/47/48/483001}{Journal of Physics A: Mathematical and Theoretical {\bf 47}, 483001}~(2014).

\bibitem{Friis:2019aa}
Nicolai Friis, Giuseppe Vitagliano, Mehul Malik, and Marcus Huber.
\newblock ``Entanglement certification from theory to experiment''.
\newblock \href{https://dx.doi.org/10.1038/s42254-018-0003-5}{Nature Reviews Physics {\bf 1}, 72--87}~(2019).

\bibitem{PhysRevLett.96.150501}
Llu\'{\i}s Masanes.
\newblock ``All bipartite entangled states are useful for information processing''.
\newblock \href{https://dx.doi.org/10.1103/PhysRevLett.96.150501}{Phys. Rev. Lett. {\bf 96}, 150501}~(2006).

\bibitem{chruscinski2014class}
Dariusz Chru{\'{s}}ci{\'{n}}ski.
\newblock ``A class of symmetric bell diagonal entanglement witnesses{\textemdash}a geometric perspective''.
\newblock \href{https://dx.doi.org/10.1088/1751-8113/47/42/424033}{Journal of Physics A: Mathematical and Theoretical {\bf 47}, 424033}~(2014).

\bibitem{PhysRevA.105.052401}
Anindita Bera, Filip~A. Wudarski, Gniewomir Sarbicki, and Dariusz Chru\ifmmode \acute{s}\else \'{s}\fi{}ci\ifmmode~\acute{n}\else \'{n}\fi{}ski.
\newblock ``Class of bell-diagonal entanglement witnesses in ${C}^{4}\ensuremath{\bigotimes}{C}^{4}$: Optimization and the spanning property''.
\newblock \href{https://dx.doi.org/10.1103/PhysRevA.105.052401}{Phys. Rev. A {\bf 105}, 052401}~(2022).

\bibitem{Choi:1975aa}
Man-Duen Choi.
\newblock ``Completely positive linear maps on complex matrices''.
\newblock \href{https://dx.doi.org/10.1016/0024-3795(75)90075-0}{Linear Algebra and its Applications {\bf 10}, 285--290}~(1975).

\bibitem{PhysRevA.84.024302}
Kil-Chan Ha and Seung-Hyeok Kye.
\newblock ``One-parameter family of indecomposable optimal entanglement witnesses arising from generalized choi maps''.
\newblock \href{https://dx.doi.org/10.1103/PhysRevA.84.024302}{Phys. Rev. A {\bf 84}, 024302}~(2011).

\bibitem{PhysRevLett.97.080501}
Heinz-Peter Breuer.
\newblock ``Optimal entanglement criterion for mixed quantum states''.
\newblock \href{https://dx.doi.org/10.1103/PhysRevLett.97.080501}{Phys. Rev. Lett. {\bf 97}, 080501}~(2006).

\bibitem{Hall_2006}
William Hall.
\newblock ``A new criterion for indecomposability of positive maps''.
\newblock \href{https://dx.doi.org/10.1088/0305-4470/39/45/020}{Journal of Physics A: Mathematical and General {\bf 39}, 14119--14131}~(2006).

\bibitem{PhysRevA.69.062311}
M.~Hein, J.~Eisert, and H.~J. Briegel.
\newblock ``Multiparty entanglement in graph states''.
\newblock \href{https://dx.doi.org/10.1103/PhysRevA.69.062311}{Phys. Rev. A {\bf 69}, 062311}~(2004).

\bibitem{Chabuda:2020aa}
Krzysztof Chabuda, Jacek Dziarmaga, Tobias~J. Osborne, and Rafa{\l} Demkowicz-Dobrza{\'n}ski.
\newblock ``Tensor-network approach for quantum metrology in many-body quantum systems''.
\newblock \href{https://dx.doi.org/10.1038/s41467-019-13735-9}{Nature Communications {\bf 11}, 250}~(2020).

\bibitem{Zhang_2021}
Zheshen Zhang and Quntao Zhuang.
\newblock ``Distributed quantum sensing''.
\newblock \href{https://dx.doi.org/10.1088/2058-9565/abd4c3}{Quantum Science and Technology {\bf 6}, 043001}~(2021).

\bibitem{Len:2022aa}
Yink~Loong Len, Tuvia Gefen, Alex Retzker, and Jan Ko{\l}ody{\'n}ski.
\newblock ``Quantum metrology with imperfect measurements''.
\newblock \href{https://dx.doi.org/10.1038/s41467-022-33563-8}{Nature Communications {\bf 13}, 6971}~(2022).

\bibitem{PhysRevLett.121.043604}
Wenchao Ge, Kurt Jacobs, Zachary Eldredge, Alexey~V. Gorshkov, and Michael Foss-Feig.
\newblock ``Distributed quantum metrology with linear networks and separable inputs''.
\newblock \href{https://dx.doi.org/10.1103/PhysRevLett.121.043604}{Phys. Rev. Lett. {\bf 121}, 043604}~(2018).

\bibitem{PhysRevA.85.062326}
M.~Zwerger, W.~D\"ur, and H.~J. Briegel.
\newblock ``Measurement-based quantum repeaters''.
\newblock \href{https://dx.doi.org/10.1103/PhysRevA.85.062326}{Phys. Rev. A {\bf 85}, 062326}~(2012).

\bibitem{Friis_2017}
Nicolai Friis, Davide Orsucci, Michalis Skotiniotis, Pavel Sekatski, Vedran Dunjko, Hans~J Briegel, and Wolfgang D{\"u}r.
\newblock ``Flexible resources for quantum metrology''.
\newblock \href{https://dx.doi.org/10.1088/1367-2630/aa7144}{New Journal of Physics {\bf 19}, 063044}~(2017).

\bibitem{PhysRevLett.120.080501}
Timothy~J. Proctor, Paul~A. Knott, and Jacob~A. Dunningham.
\newblock ``Multiparameter estimation in networked quantum sensors''.
\newblock \href{https://dx.doi.org/10.1103/PhysRevLett.120.080501}{Phys. Rev. Lett. {\bf 120}, 080501}~(2018).

\bibitem{Amselem:2009aa}
Elias Amselem and Mohamed Bourennane.
\newblock ``Experimental four-qubit bound entanglement''.
\newblock \href{https://dx.doi.org/10.1038/nphys1372}{Nature Physics {\bf 5}, 748--752}~(2009).

\bibitem{PhysRevLett.105.130501}
Jonathan Lavoie, Rainer Kaltenbaek, Marco Piani, and Kevin~J. Resch.
\newblock ``Experimental bound entanglement in a four-photon state''.
\newblock \href{https://dx.doi.org/10.1103/PhysRevLett.105.130501}{Phys. Rev. Lett. {\bf 105}, 130501}~(2010).

\bibitem{PhysRevLett.109.040501}
Fumihiro Kaneda, Ryosuke Shimizu, Satoshi Ishizaka, Yasuyoshi Mitsumori, Hideo Kosaka, and Keiichi Edamatsu.
\newblock ``Experimental activation of bound entanglement''.
\newblock \href{https://dx.doi.org/10.1103/PhysRevLett.109.040501}{Phys. Rev. Lett. {\bf 109}, 040501}~(2012).

\bibitem{ma2012experimental}
Xiao-song Ma, Stefan Zotter, Johannes Kofler, Rupert Ursin, Thomas Jennewein, {\v C}aslav Brukner, and Anton Zeilinger.
\newblock ``Experimental delayed-choice entanglement swapping''.
\newblock \href{https://dx.doi.org/10.1038/nphys2294}{Nature Physics {\bf 8}, 479--484}~(2012).

\bibitem{Dur_2007}
W~D{\"u}r and H~J Briegel.
\newblock ``Entanglement purification and quantum error correction''.
\newblock \href{https://dx.doi.org/10.1088/0034-4885/70/8/R03}{Reports on Progress in Physics {\bf 70}, 1381}~(2007).

\bibitem{l43p-py55}
Jie Zhou, Chuanlong Ma, Yifang Xu, Weizhou Cai, Hongwei Huang, Lida Sun, Guangming Xue, Ziyue Hua, Haifeng Yu, Weiting Wang, Chang-Ling Zou, and Luyan Sun.
\newblock ``Experimental demonstration of entanglement pumping with bosonic logical qubits''.
\newblock \href{https://dx.doi.org/10.1103/l43p-py55}{Phys. Rev. Lett. {\bf 136}, 110801}~(2026).

\bibitem{PhysRevA.76.030305}
Otfried G\"uhne, Chao-Yang Lu, Wei-Bo Gao, and Jian-Wei Pan.
\newblock ``Toolbox for entanglement detection and fidelity estimation''.
\newblock \href{https://dx.doi.org/10.1103/PhysRevA.76.030305}{Phys. Rev. A {\bf 76}, 030305}~(2007).

\bibitem{Haffner:2005aa}
H.~H{\"a}ffner, W.~H{\"a}nsel, C.~F. Roos, J.~Benhelm, D.~Chek-al kar, M.~Chwalla, T.~K{\"o}rber, U.~D. Rapol, M.~Riebe, P.~O. Schmidt, C.~Becher, O.~G{\"u}hne, W.~D{\"u}r, and R.~Blatt.
\newblock ``Scalable multiparticle entanglement of trapped ions''.
\newblock \href{https://dx.doi.org/10.1038/nature04279}{Nature {\bf 438}, 643--646}~(2005).

\bibitem{PhysRevLett.95.210502}
Nikolai Kiesel, Christian Schmid, Ulrich Weber, G\'eza T\'oth, Otfried G\"uhne, Rupert Ursin, and Harald Weinfurter.
\newblock ``Experimental analysis of a four-qubit photon cluster state''.
\newblock \href{https://dx.doi.org/10.1103/PhysRevLett.95.210502}{Phys. Rev. Lett. {\bf 95}, 210502}~(2005).

\bibitem{Lu:2007aa}
Chao-Yang Lu, Xiao-Qi Zhou, Otfried G{\"u}hne, Wei-Bo Gao, Jin Zhang, Zhen-Sheng Yuan, Alexander Goebel, Tao Yang, and Jian-Wei Pan.
\newblock ``Experimental entanglement of six photons in graph states''.
\newblock \href{https://dx.doi.org/10.1038/nphys507}{Nature Physics {\bf 3}, 91--95}~(2007).

\bibitem{PhysRevA.63.032306}
John~A. Smolin.
\newblock ``Four-party unlockable bound entangled state''.
\newblock \href{https://dx.doi.org/10.1103/PhysRevA.63.032306}{Phys. Rev. A {\bf 63}, 032306}~(2001).

\bibitem{Stormer:1963aa}
Erling St{\o}rmer.
\newblock ``Positive linear maps of operator algebras''.
\newblock \href{https://dx.doi.org/10.1007/BF02391860}{Acta Mathematica {\bf 110}, 233--278}~(1963).

\bibitem{10.1145/780542.780545}
Leonid Gurvits.
\newblock ``Classical deterministic complexity of edmonds' problem and quantum entanglement''.
\newblock In Proceedings of the Thirty-Fifth Annual ACM Symposium on Theory of Computing.
\newblock \href{https://dx.doi.org/10.1145/780542.780545}{Pages 10--19}.
\newblock STOC '03New York, NY, USA~(2003). Association for Computing Machinery.

\bibitem{PhysRevA.61.062313}
W.~D\"ur, J.~I. Cirac, M.~Lewenstein, and D.~Bru\ss{}.
\newblock ``Distillability and partial transposition in bipartite systems''.
\newblock \href{https://dx.doi.org/10.1103/PhysRevA.61.062313}{Phys. Rev. A {\bf 61}, 062313}~(2000).

\bibitem{PhysRevLett.77.1413}
Asher Peres.
\newblock ``Separability criterion for density matrices''.
\newblock \href{https://dx.doi.org/10.1103/PhysRevLett.77.1413}{Phys. Rev. Lett. {\bf 77}, 1413--1415}~(1996).

\bibitem{PhysRevLett.88.247901}
Karl Gerd~H. Vollbrecht and Michael~M. Wolf.
\newblock ``Activating distillation with an infinitesimal amount of bound entanglement''.
\newblock \href{https://dx.doi.org/10.1103/PhysRevLett.88.247901}{Phys. Rev. Lett. {\bf 88}, 247901}~(2002).

\bibitem{doi:10.1142/S1230161213500066}
Dariusz Chru{\'s}ci\'{n}ski and Gniewomir Sarbicki.
\newblock ``Optimal entanglement witnesses for two qutrits''.
\newblock \href{https://dx.doi.org/10.1142/S1230161213500066}{Open Systems \& Information Dynamics {\bf 20}, 1350006}~(2013).

\bibitem{PhysRevA.59.4206}
Micha\l{} Horodecki and Pawe\l{} Horodecki.
\newblock ``Reduction criterion of separability and limits for a class of distillation protocols''.
\newblock \href{https://dx.doi.org/10.1103/PhysRevA.59.4206}{Phys. Rev. A {\bf 59}, 4206--4216}~(1999).

\bibitem{PhysRevA.62.062314}
W.~D\"ur, G.~Vidal, and J.~I. Cirac.
\newblock ``Three qubits can be entangled in two inequivalent ways''.
\newblock \href{https://dx.doi.org/10.1103/PhysRevA.62.062314}{Phys. Rev. A {\bf 62}, 062314}~(2000).

\bibitem{PhysRevA.61.062312}
David~P. DiVincenzo, Peter~W. Shor, John~A. Smolin, Barbara~M. Terhal, and Ashish~V. Thapliyal.
\newblock ``Evidence for bound entangled states with negative partial transpose''.
\newblock \href{https://dx.doi.org/10.1103/PhysRevA.61.062312}{Phys. Rev. A {\bf 61}, 062312}~(2000).

\bibitem{Cavalcanti2013EntanglementSteering}
Eric~G. Cavalcanti, Michael J.~W. Hall, and Howard~M. Wiseman.
\newblock ``Entanglement verification and steering when alice and bob cannot be trusted''.
\newblock \href{https://dx.doi.org/10.1103/PhysRevA.87.032306}{Physical Review A {\bf 87}, 032306}~(2013).

\bibitem{zhao2020experimental}
Yuan-Yuan Zhao, Huan-Yu Ku, Shin-Liang Chen, Hong-Bin Chen, Franco Nori, Guo-Yong Xiang, Chuan-Feng Li, Guang-Can Guo, and Yueh-Nan Chen.
\newblock ``Experimental demonstration of measurement-device-independent measure of quantum steering''.
\newblock \href{https://dx.doi.org/10.1038/s41534-020-00307-9}{npj Quantum Information {\bf 6}, 77}~(2020).

\bibitem{Rosset2018QuantumMemories}
Denis Rosset, Francesco Buscemi, and Yeong-Cherng Liang.
\newblock ``Resource theory of quantum memories and their faithful verification with minimal assumptions''.
\newblock \href{https://dx.doi.org/10.1103/PhysRevX.8.021033}{Physical Review X {\bf 8}, 021033}~(2018).

\bibitem{PhysRevA.84.032310}
Bastian Jungnitsch, Tobias Moroder, and Otfried G\"uhne.
\newblock ``Entanglement witnesses for graph states: General theory and examples''.
\newblock \href{https://dx.doi.org/10.1103/PhysRevA.84.032310}{Phys. Rev. A {\bf 84}, 032310}~(2011).

\end{thebibliography}

\onecolumn
\appendix

\section*{Appendix}

\section{Entanglement detection by state preparation for two-qubit states}
\label{app1}

We here reproduce a network state for two-qubit EWs. Let $|\phi^{\pm}\rangle = (|00\rangle\pm |11\rangle )/\sqrt{2}$ and $|\psi^{\pm}\rangle = (|01\rangle\pm |10\rangle )/\sqrt{2}$ denote four Bell states. We show how to construct a network state for an EW 
\bea
W = |\phi^+\rangle \langle \phi^+|^{\Gamma} = \frac{1}{2}\mathbbm{I} - |\psi^-\rangle \langle \psi^-|. \nonumber
\eea
One may find a decomposition of the above EW in the following way
\bea
W = \frac{1}{2} (\mathbbm{I} -  |\psi^-\rangle \langle \psi^-|) -\frac{1}{2} |\psi^-\rangle \langle \psi^-|, \nonumber
\eea
where two non-negative operators are obtained as $ |\psi^-\rangle \langle \psi^-|$ and $(\mathbbm{I} -  |\psi^-\rangle \langle \psi^-|)/2$. A network state may be written as
\bea
N_{23} &=& c_1 |\psi^-\rangle_{A_2 B_2 } \langle \psi^-| \otimes \Pi_{A_3B_3}(1) + \nonumber\\
&& \frac{c_2}{3} (\mathbbm{I} - |\psi^-\rangle_{A_2 B_2 } \langle \psi^-| ) \otimes \Pi_{A_3B_3}(2), \nonumber
\eea
for some positive constants $c_1,~c_2=1-c_1>0$ and non-negative normalized operators $\Pi(1),~ \Pi(2)\geq 0$. To realize entanglement detection of a state of interest $\rho$ using an entanglement witness $W$, one may seek $N_{23}$ that satisfy 
\bea
\tr[\rho_1 W_1] =  16~ \tr  [\rho_1\otimes N_{23} (\eta \mathbbm{1} - | \phi^+\rangle_{A_3B_3} \langle \phi^+|) \otimes P^{(12)}  ], \nonumber 
\eea
with $\eta = 1/2$, since all two-qubit entangled states are distillable. The goal is now to find the parameters $c_1,~c_2,~ \Pi(1)$ and $\Pi(2)$ that satisfy the relation in the above. The left-hand-side (lhs) is given by
\bea
\mathrm{lhs}= \frac{1}{2} - \langle \psi^-| \rho|\psi^-\rangle, \nonumber
\eea
and the right-hand-side (rhs) by
\bea 
\mathrm{rhs}&=& 2 \tr [\rho_{2}^T N_{23}] - 4 \tr [\rho_{2}^T \otimes |\psi^-\rangle \langle \psi^-|  N_{23} ]    \nonumber \\
&=& c_2( \frac{2}{3}  - \frac{4}{3} \langle \phi^+| \Pi (2) |\phi^+\rangle )  + L \langle \psi^-| \rho |\psi^-\rangle, \nonumber
\eea
where 
\bea
L &=& c_1 (2-4 \langle \phi^+| \Pi(1) | \phi^+\rangle ) +  \nonumber \\
 && c_2 ( - \frac{2}{3}  + \frac{4}{3} \langle \phi^+| \Pi(2) | \phi^+\rangle ). \nonumber
\eea
From the lhs and the rhs, one can find that
\bea
c_2( \frac{2}{3}  - \frac{4}{3} \langle \phi^+| \Pi (1) |\phi^+\rangle )= \frac{1}{2}~\mathrm{and} ~L=-1, \nonumber
\eea
from which 
\bea
&& c_1 (2-4 \langle \phi^+| \Pi(1) | \phi^+\rangle ) = - \frac{1}{2} \nonumber\\
& \iff &c_1 = \frac{1}{8 \langle \phi^+ | \Pi(1) |\phi^+\rangle -4 } >0. \nonumber
\eea
For convenience, we choose $\Pi(1) = |\phi^+\rangle \langle \phi^+|$ although it is not a unique choice. It follows that $c_1=1/4$ and $c_2=3/4$. The consequence is that $\langle \phi^+| \Pi(2) |\phi^+\rangle = 0$. Thus, we have 
\bea
\Pi(2) = \frac{1}{3} ( \mathbbm{I} -|\phi^+\rangle \langle \phi^+| ).\nonumber
\eea
All these conclude a network state
\bea
N_{23} &=& \frac{1}{4} |\psi^-\rangle_{A_2B_2} \langle \psi^-| \otimes |\phi^+\rangle_{A_3B_3} \langle \phi^+| + \nonumber \\
&&\frac{1}{12} (\mathbbm{1} - |\psi^-\rangle_{A_2B_2} \langle \psi^-|) \otimes (\mathbbm{1} - |\phi^+\rangle_{A_3B_3} \langle \phi^+|). \nonumber
\eea
Note that a network state for an EW is not unique.

\section{Network states for high-dimensional EWs}
\label{app2}

\subsection{The framework}

Recall that for a given EW $W$, we are looking for a network state $N_{23} = N^{(A_2 B_2 A_3 B_3)}$, which are separable in $A_2 B_2: A_3 B_3$, satisfying the following condition:
\bea
W_2^T \propto \tr_3[N_{23} (\eta \mathbbm{1} - P_{00})_3], 
\eea
for some $\eta \in [\frac{1}{d}, 1)$.
It is easy to see that the following relation holds:
\bea
\tr[\rho_1 W_1] &=& d^2 \tr[\rho_1 \otimes W_2^T P^{(12)}]
\\ &\propto& \tr[\rho_1 \otimes N_{23} ~ P^{(12)} \otimes (\eta \mathbbm{1} - P_{00})_3], \label{prop}
\eea
where $P^{(12)} = P_{00}^{(A_1 A_2)} \otimes P_{00}^{(B_1 B_2)}$ denotes the Bell measurements on both sides.
The scheme can be understood as follows. First, a filtering operation $\Lambda^{(1 \to 3)}$ teleports the state $\rho_1$ to $A_3 B_3$, leaving a result state $\Lambda(\rho)$:
\bea
\Lambda^{(1 \to 3)} (\rho) = \frac{\tr_{12}[\rho_1 \otimes N_{23} P^{(12)}] }{\tr[\rho_1 \otimes N_{23} P^{(12)}] }. \label{filtering}
\eea
Then the singlet fraction, or the overlap with the Bell state $P_{00}$, of the resulting state $\Lambda(\rho)$ is checked whether it is higher than $\eta$ or not. 

The singlet fraction can also be estimated with a fixed measurement on individual quantum systems. The main idea is to place unitary interactions before a measurement. A $d$-dimensional Hadamard gate and a $d$-dimensional CNOT gate may be obtained as,
\begin{align*}
H &= \sum_{j,k=0}^{d-1} e^{2 \pi ijk/d} \ket{j}\bra{k},~ \mathrm{and}
\\ U_{CNOT} &= \sum_{j=0}^{d-1} \ket{j}\bra{j} \otimes \sum_{k=0}^{d-1} \ket{k+j}\bra{k}.    
\end{align*}
Note that a maximally entangled state can be generated,  $\ket{\phi_{00}} = U_{CNOT}(H \otimes \I) \ket{00}$. Then, instead of a joint measurement, one can first apply $(H^\dagger \otimes \I) U_{CNOT}^\dagger $ to a resulting state $\Lambda^{(1 \to 3)} (\rho)$ and then perform a measurement in the computational basis. The probability of having outcomes $00$ gives the singlet fraction $\langle \phi_{00} | \Lambda(\rho) | \phi_{00} \rangle$.
It holds that $\langle \phi_{00} | \Lambda(\rho) | \phi_{00} \rangle > \eta$ if and only if $\tr[\rho W] < 0$, which certifies that a state $\rho$ given in the beginning is entangled. \\

\subsection{Decomposable EW} \label{app:decEW}

Consider a decomposable EW $W = Q^\Gamma$ for $Q\geq 0$ and $\tr[Q]=1$. Let $\lambda:=\max_{i}|\lambda_i|$ where $\{\lambda_i\} $ are eigenvalues of $W$. Then, a network state for the EW is obtained as
\bea
N_{23}^{\text{(dec)}} &=& c_1\left( \frac{\lambda\mathbbm{1} - Q^\Gamma }{\lambda d^2 - 1 }\right)^{(2)} \otimes P_{00}^{(3)} \nonumber\\
&&+ c_2 \left( \frac{ \lambda \mathbbm{1} +  Q^{\Gamma} }{\lambda d^2 + 1}\right)^{(2)} \otimes \left( \frac{\mathbbm{1} - P_{00}}{d^2 - 1}\right)^{(3)},~~ \label{decom NS}
\eea
with the threshold value $\eta = \frac{1}{d}$, 
\bea
c_1 = \frac{d^2 \lambda - 1}{d^3 \lambda  + d -2} ~~\mathrm{and}~ ~ c_2= \frac{(d-1)(d^2 \lambda +1)}{d^3 \lambda  +d-2}. \nonumber
\eea
The superscript $(j)$ stands for the composite space $A_j B_j$ for $j \in \{1,2,3\}$.
From the equation
\bea
\tr_3[ N_{23 } (\frac{1}{d}\mathbbm{1}-P_{00})_3] = \frac{2(d-1)}{d(d^3 \lambda + d -2)} Q^\Gamma, \nonumber
\eea
it holds that
\bea
& \tr[\rho W] = k ~ \tr[\rho_1 \otimes N_{23} ~ P^{(12)} \otimes (\frac{1}{d}\mathbbm{1}-P_{00})_3], & \nonumber
\\ & \bracket{\phi_{00}}{\Lambda(\rho)}{\phi_{00}} > \frac{1}{d }\iff \tr[\rho W] <0, &
\eea
where $k=\frac{d(d^3 \lambda + d -2)}{2(d-1)}$.

Hence, the partial transpose criteria can be realized by preparing a network state in Eq. (\ref{decom NS}). In particular, a network state for the decomposable EW 
\bea
W = P_{00}^{\Gamma} = \frac{\mathbbm{F}}{d},
\eea
which is proportional to the flip operator $\mathbbm{F}$ that detects entangled Werner states, can be found as 
\bea
N_{23}^{\text{(flip)}} &=& \frac{1}{d+2} \left( \frac{\mathbbm{1}-\mathbbm{F}}{d^2-d} \right)^{(2)} \otimes P_{00}^{(3)}  \nonumber
\\ && + \frac{d+1}{d+2} \left( \frac{\mathbbm{1}+\mathbbm{F}}{d^2+d} \right)^{(2)} \otimes \left( \frac{\mathbbm{1}-P_{00}}{d^2-1} \right)^{(3)}.~~ \label{flip NS}
\eea
Note that this network state is invariant under $U_{A_2} \otimes U_{B_2} \otimes V_{A_3} \otimes V^*_{B_3}$ for any unitary operation $U, V$. This network state is positive under the partial transpose $A_2 A_3: B_2 B_3$, so it is undistillable. This state (\ref{flip NS}) has been used in the activation of non-PPT entangled states and proved to be PPT in \cite{PhysRevLett.88.247901}.

\subsection{ Bell-diagonal EW} \label{app:BDEW}
The next examples are Bell-diagonal witnesses $W(\Vec{\lambda})$, which are decomposable or non-decomposable depending on the parameter $\Vec{\lambda} = (\lambda_0, \ldots, \lambda_{d-1})$:
\bea
W(\Vec{\lambda}) = \sum_{s=0}^{d-1} \lambda_s \Pi_s - P_{00}, \label{eq:Wa2}
\eea
where $\lambda_s \ge 0 ~\forall s$ and $\sum_{s=0}^{d-1} \lambda_s = 1$. 
Note also that $W[\Vec{\lambda}]$ in Eq. (\ref{eq:Wa2}) is an EW if a vector $\Vec{\lambda}$ satisfies the cyclic inequalities in the following, 
\bea
\sum_{j=0}^{d-1} \frac{t_{j}^2}{\sum_{s=0}^{d-1} \lambda_s t_{j+s}^2} \le d, \nonumber
\eea
for all $t_0, \ldots, t_d \ge 0$.
The value $\lambda_0$ is critical in implementing this witness with state preparation, as shown below. To construct a network state for a Bell-diagonal EW, we use paired Bell-diagonal (PBD) states:
\bea
N_{23}^{\textrm{(PBD)}} (\Vec{\lambda}) = \sum_{s=0}^{d-1} \lambda_s \frac{1}{d} \sum_{t=0}^{d-1} P_{st}^{(2)} \otimes P_{st}^{(3)},
\eea
with the threshold value $\eta = \lambda_0$. 
Then the following holds:
\bea
& \tr[\rho W(\Vec{\lambda})] = k ~\tr[\rho_1 \otimes N_{23}^{\textrm{(PBD)}}(\Vec{\lambda})~ P^{(12)} \otimes (\lambda_0 \mathbbm{1} - P_{00})_3 ], & \nonumber
\\ & \bracket{\phi_{00}}{\Lambda(\rho)}{\phi_{00}} > \lambda_0 \iff \tr[\rho W(\Vec{\lambda})] <0. \nonumber &
\eea
where $k = d^3 / \lambda_0$. It is possible to achieve $\bracket{\phi_{00}}{\Lambda(\sigma)}{\phi_{00}} = \lambda_0$ with a separable state $\sigma$:
\bea
\Lambda^{(1 \to 3)} (\sigma) &=& \frac{\tr_{12}[\sigma_1 \otimes N_{23} P^{(12)}] }{\tr[\sigma_1 \otimes N_{23} P^{(12)}] }, \nonumber
\\ \textrm{where } \sigma &=& \frac{1}{d} P_{00} + \frac{1}{d} \sum_{s=1}^{d-1} \frac{\Pi_s}{d}. \nonumber
\eea
In particular, EWs from a reduction map and  the Choi map can be written in the form of Bell-diagonal witness and can be implemented by preparing the corresponding PBD state.
A decomposable EW from the reduction map corresponds to a Bell-diagonal EW with $\Vec{\lambda} = (\frac{1}{d}, \ldots, \frac{1}{d})$:
\bea
W_{\textrm{red}} &=& \sum_{s=0}^{d-1} \frac{1}{d} \Pi_s - P_{00} = \frac{1}{d} \mathbbm{1} - P_{00} \label{red},
\\ N_{23}^{\textrm{(red)}} &=& \frac{1}{d^2} \sum_{s=0}^{d-1} \sum_{t=0}^{d-1} P_{st}^{(2)} \otimes P_{st}^{(3)},
\eea
with the threshold value $\eta = \frac{1}{d}$.

The PBD state $N_{23}^{\textrm{(red)}}$ can be seen as a direct generalization of Smolin state \cite{PhysRevA.63.032306} into $d$-dimension. In $d=2$, Smolin state is PPT in $A_2 A_3:B_2 B_3$. However, the state $N_{23}^{\textrm{(red)}}$ is not PPT in higher dimensions $d \ge 3$ in general. Smolin state in two-dimension is undistillable, but the distillability of $N_{23}^{\textrm{(red)}}$ in higher dimensions is unknown.

Although it can be generalized into higher dimensions \cite{PhysRevA.84.024302, doi:10.1142/S1230161213500066}, the nondecomposable witness from Choi map \cite{Choi:1975aa} is defined in $d=3$ and corresponds to a Bell-diagonal witness with $\Vec{\lambda} = (\frac{2}{3}, \frac{1}{3}, 0)$:
\bea
W_{\textrm{Choi}} &=& \frac{2}{3} \Pi_0 + \frac{1}{3} \Pi_1 - P_{00},
\\ N_{23}^{\textrm{(Choi)}} &=& \frac{2}{9} \sum_{t=0}^{2} P_{0,t}^{(2)} \otimes P_{0,t}^{(3)} + \frac{1}{9} \sum_{t=0}^{2} P_{1,t}^{(2)} \otimes P_{1,t}^{(3)},~~~~~~
\eea
with the threshold value $\eta = \frac{2}{3}$.

\subsection{  Breuer-Hall EW} \label{app:BH}

The Breuer-Hall map \cite{PhysRevLett.97.080501, Hall_2006} finds an EW,
\bea
W_{\textrm{BH}} = \frac{1}{d} (\mathbbm{1}- \mathbbm{F}') - P_{00} ,
\eea
where $\mathbbm{F}' = (\mathbbm{1} \otimes U) \mathbbm{F} (\mathbbm{1} \otimes U^\dagger)$ for any skew-symmetric unitary operator $U$ such that $UU^\dagger = \mathbbm{1}$ and $U^T = -U$. In even dimensions $d=2n$, one can set $U$ as
\bea
U = \Oplus_{i=1}^{n} 
\begin{bmatrix}
0 & 1 \\
-1 & 0 
\end{bmatrix}, \nonumber
\eea
then it acts as $U \ket{i} = (-1)^{j} \ket{j}$ where $j = i+1$ for even $i$ and $j = i-1$ for odd $i$.

One can think of $W_{\textrm{BH}}$ as a combination of reduction EW (\ref{red}) and the flip operator from reduction map with additional $U$.
A network state for the BH EW is as follows,
\bea
N_{23}^{(\textrm{BH})}  
& = & c_0 \frac{1}{d^2} \sum_{s=0}^{d-1} \sum_{t=0}^{d-1} P_{st}^{(2)}  \otimes P_{st}^{(3)} \nonumber \\
&& + c_1 \left( \frac{\mathbbm{1} + \mathbbm{F}' }{d^2+d} \right)^{(2)} \otimes P_{00}^{(3)}   \nonumber \\
&& + c_2 \left( \frac{\mathbbm{1} - \mathbbm{F}' }{d^2-d} \right)^{(2)} \otimes \left( \frac{\mathbbm{1} - P_{00}}{d^2-1} \right)^{(3)},~~~~ \label{eq:nbh} 
\eea
with the threshold value $\eta = \frac{1}{d}$, where $c_0 = \frac{2d^2-2d}{3d^2 -3d +2}$, $c_1= \frac{d+1}{3d^2-3d+2}$, and $c_2 = \frac{d^2-2d+1}{3d^2-3d+2}$. One can show that
\bea
 \tr_3[ N_{23}^{(\mathrm{BH})}~ (\eta \mathbbm{1} - P_{00})_3] = \frac{c_0}{d^2} W_{\mathrm{BH}}^T,
\eea
which leads to
\bea
& \tr[\rho W_{\textrm{BH}}] = k ~\tr[\rho_1 \otimes N_{23}^{\textrm{(BH)}}~ P^{(12)} \otimes (\eta \mathbbm{1} - P_{00})_3 ], & \nonumber
\\ & \bracket{\phi_{00}}{\Lambda(\rho)}{\phi_{00}} > \frac{1}{d} \iff \tr[\rho W_{\textrm{BH}}] <0, \nonumber &
\eea
where $k =  d^4 / c_0$.


\begin{figure}[t]
\centering
\resizebox{0.4\linewidth}{!}{
\begin{tikzpicture}[scale=0.5]
    \definecolor{customBlue}{rgb}{0.2, 0.4, 1}
    \definecolor{customGreen}{rgb}{0.2, 1, 0.2}

    \fill[lightgray, rounded corners=0.15cm] (2-0.35,3+1-0.35) rectangle (2+0.35,6+1+0.35);
    \fill[lightgray, rounded corners=0.15cm] (6-0.35,3+1-0.35) rectangle (6+0.35,6+1+0.35);
    \fill[lightgray, rounded corners=0.15cm] (10-0.35,3+1-0.35) rectangle (10+0.35,6+1+0.35);

    \draw[customGreen, ultra thick] (0,0+0) -- (4,0+0) -- (8,0+0);  
    \draw[customGreen, ultra thick] (2,0+1) -- (6,0+1) -- (10,0+1); 
    \draw[customGreen, ultra thick] (0,0+0) -- (2,0+1);  
    \draw[customGreen, ultra thick] (4,0+0) -- (6,0+1);
    \draw[customGreen, ultra thick] (8,0+0) -- (10,0+1);

    \draw[customGreen, ultra thick] (2,3+1) -- (6,3+1) -- (10,3+1);

    \fill[lightgray, rounded corners=0.15cm] (0-0.35,3+0-0.35) rectangle (0+0.35,6+0+0.35);
    \fill[lightgray, rounded corners=0.15cm] (4-0.35,3+0-0.35) rectangle (4+0.35,6+0+0.35);
    \fill[lightgray, rounded corners=0.15cm] (8-0.35,3+0-0.35) rectangle (8+0.35,6+0+0.35);

    \draw[customGreen, ultra thick] (0,3+0) -- (4,3+0) -- (8,3+0);
    \draw[customGreen, ultra thick] (0,3+0) -- (2,3+1);
    \draw[customGreen, ultra thick] (4,3+0) -- (6,3+1);
    \draw[customGreen, ultra thick] (8,3+0) -- (10,3+1);
    
    \draw[customBlue, ultra thick] (0,6+0) -- (4,6+0) -- (8,6+0);
    \draw[customBlue, ultra thick] (2,6+1) -- (6,6+1) -- (10,6+1);
    \draw[customBlue, ultra thick] (0,6+0) -- (2,6+1);
    \draw[customBlue, ultra thick] (4,6+0) -- (6,6+1);
    \draw[customBlue, ultra thick] (8,6+0) -- (10,6+1);

    \fill[customGreen] (0,0+0) circle (0.3);
    \fill[customGreen] (4,0+0) circle (0.3);
    \fill[customGreen] (8,0+0) circle (0.3);
    \fill[customGreen] (2,0+1) circle (0.3);
    \fill[customGreen] (6,0+1) circle (0.3);
    \fill[customGreen] (10,0+1) circle (0.3);

    \fill[customGreen] (0,3+0) circle (0.3);
    \fill[customGreen] (4,3+0) circle (0.3);
    \fill[customGreen] (8,3+0) circle (0.3);
    \fill[customGreen] (2,3+1) circle (0.3);
    \fill[customGreen] (6,3+1) circle (0.3);
    \fill[customGreen] (10,3+1) circle (0.3);

    \fill[customBlue] (0,6+0) circle (0.3);
    \fill[customBlue] (4,6+0) circle (0.3);
    \fill[customBlue] (8,6+0) circle (0.3);
    \fill[customBlue] (2,6+1) circle (0.3);
    \fill[customBlue] (6,6+1) circle (0.3);
    \fill[customBlue] (10,6+1) circle (0.3);

    \draw[customGreen, ultra thick, dotted] (0,0+0) -- (0,3+0);
    \draw[customGreen, ultra thick, dotted] (4,0+0) -- (4,3+0);
    \draw[customGreen, ultra thick, dotted] (8,0+0) -- (8,3+0);
    \draw[customGreen, ultra thick, dotted] (2,0+1) -- (2,3+1);
    \draw[customGreen, ultra thick, dotted] (6,0+1) -- (6,3+1);
    \draw[customGreen, ultra thick, dotted] (10,0+1) -- (10,3+1);

    \node at (10,0+0.25) [text=customGreen] {\huge \(\cdot\)\(\cdot\)\(\cdot\)};
    \node[fill=white, inner sep=0pt] at (10,3+0.25) [text=customGreen] {\huge \(\cdot\)\(\cdot\)\(\cdot\)};
    \node at (10,6+0.25) [text=customBlue] {\huge \(\cdot\)\(\cdot\)\(\cdot\)};

    \node[fill=white, inner sep=0pt] at (1,1.75) {$N_{23}$};
    \node at (0.65,6.85) {$\rho_1$};
\end{tikzpicture}
}
\caption{ A graph state can be detected by preparing a network state, Bell measurements, and a fixed measurement. }
\label{fig:figm}
\end{figure}

\section{Entanglement witnesses for graph states} \label{app:graph}
\label{app3}
A graph $G = (V ,E)$ is defined by a set of vertices $V$ and a set of edges $E$:
\bea
V &=& \{1,\ldots,n\}, ~ \text{where $n$ is the number of vertices,} \nonumber
\\ E &=& \{(i,j)| i,j \in V, i<j, \text{Vertex $i$ and $j$ are connected.}\} \nonumber
\eea
Also define the neighborhood of vertex $i$: $E_i = \{j \in V| (i,j) \in E \text{ or } (j,i) \in E\}$, which is a set of vertices connected to vertex $i$.

\begin{figure}[t]
\centering
\includegraphics[width=7cm]{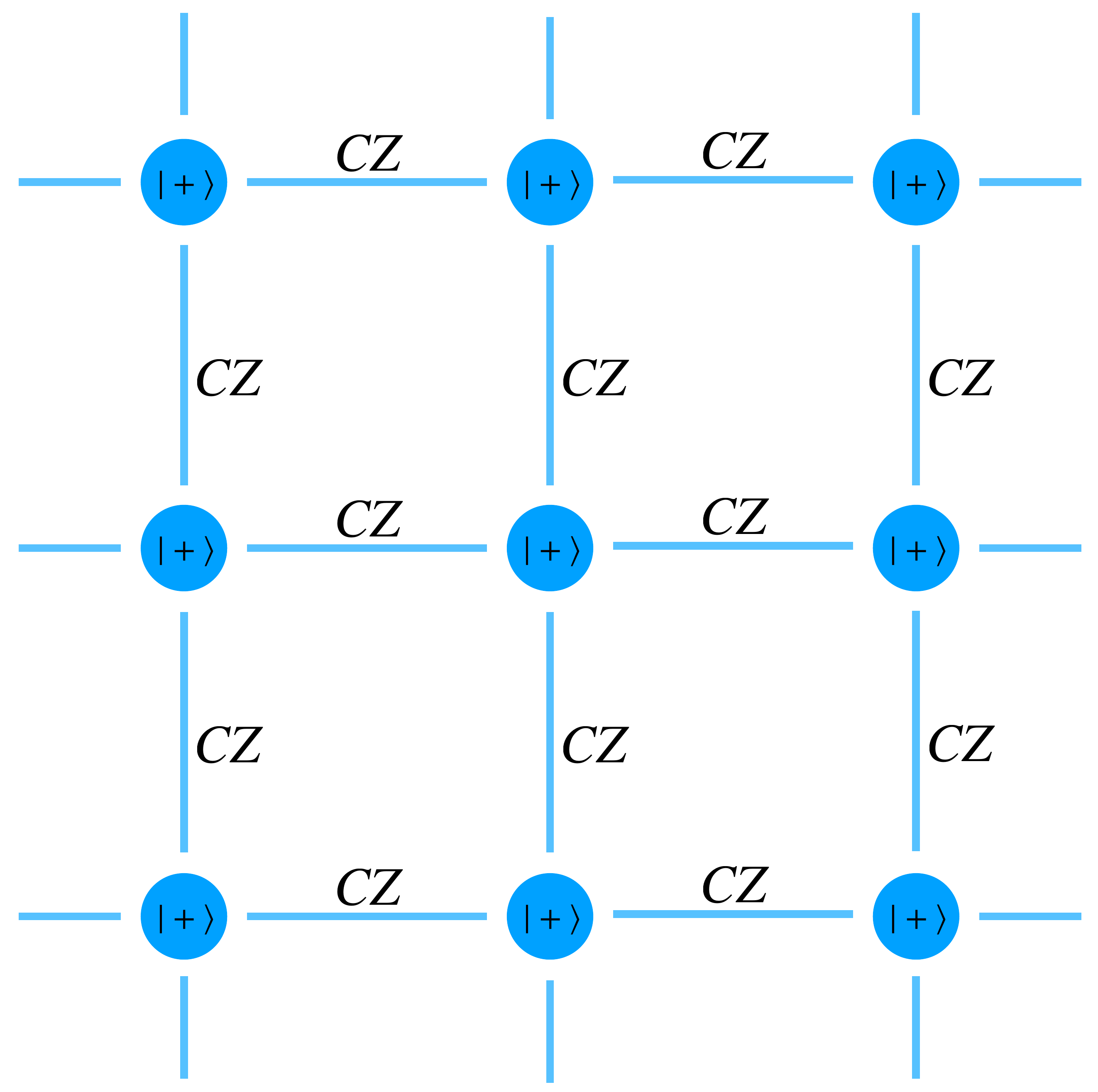}
\caption{A graph consisting of a set of vertices and a set of edges uniquely defines the graph state. The graph state can be obtained by preparing $\ket{+} = (\ket{0}+\ket{1})/\sqrt{2}$ at all vertices in $V$, and applying the controlled-Z ($CZ$) gats to all edges in $E$.}
\label{fig:graph}
\end{figure}

The generators $g_i$ and the projectors $\gamma_i$ of a graph state determined by the graph $G$ are given by
\bea
g_i &=& X_i \prod_{j \in E_i} Z_j, \nonumber
\\ \gamma_i^{(x_i)} &=& \frac{\mathbbm{1}+(-1)^{x_i} g_i}{2}.
\eea
Note that the eigenvalue of $g_i$ is either 1 or -1, and the eigenvalue of $\gamma_i$ is either 1 or 0. Now we define an orthonormal graph state basis consisting of $2^n$ states:
\bea
\ket{\Vec{x}}\bra{\Vec{x}}_G = \prod_{i=1}^{n} \gamma_i^{(x_i)}, ~~ \text{where } \Vec{x} = (x_1, \ldots, x_n) \in \{0,1\}^n. \nonumber
\eea
The states $\ket{\Vec{x}}_G$ are the eigenstates of the generators and projectors: 
\bea
g_i \ket{\Vec{x}}_G &=& (-1)^{x_i} \ket{\Vec{x}}_G, \\
\gamma_i^{(k)} \ket{\Vec{x}}_G &=&
\begin{cases}
\ket{\Vec{x}}_G & \mbox{if}\; x_i = k,\\
0 & \mbox{if}\; x_i \ne k.
\end{cases}
\eea
A graph state, denoted by $\ket{\Vec{0}}\bra{\Vec{0}}_G$, corresponds to an eigenstate with eigenvalue $+1$ for all generators $g_i$ ($i \in \{1, \ldots, n\}$). A state can be obtained by preparing $\ket{+} = (\ket{0}+\ket{1})/\sqrt{2}$ placed at vertices in $V$, and applying the controlled-Z ($CZ$) gate to all edges in $E$, see Fig. \ref{fig:graph}.

\begin{figure}[t]
\centering
\includegraphics[width=8cm]{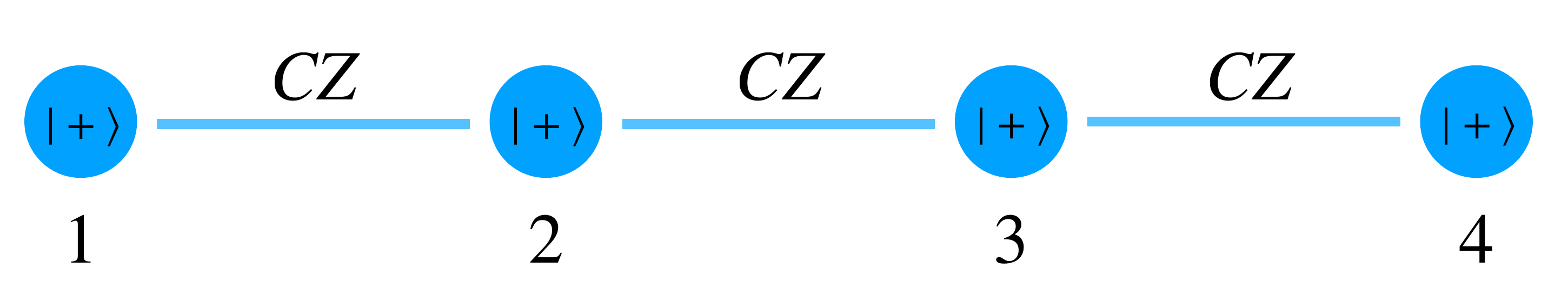}
\caption{The four-qubit linear cluster state on the graph $G_{Cl4}$. One can obatin the graph state $\ketbra{0000}_G$ by preparing four $\ket{+}$ states and applying the controlled-Z gate to three edges.}
\label{fig:Cl4}
\end{figure}

For instance, the four-qubit linear cluster state can be defined by a graph $G_{Cl4} = (V_4,E_{Cl4})$, where $V_4 = \{1,2,3,4\}$ and $E_{Cl4} = \{(1,2),(2,3),(3,4)\}$, see Fig. \ref{fig:Cl4}. 
The generators define $16$ graph states $\ket{0000}\bra{0000}_G, \ldots,$ and $\ket{1111}\bra{1111}_G$. The graph state $\ket{0000}\bra{0000}_G$ is detected by an EW in the following,
\bea
W_{Cl4} = \frac{1}{2} \sum_{\Vec{x} \in S} \ket{\Vec{x}}\bra{\Vec{x}}_G - \ketbra{0000}_G,
\eea
where $S = \{$0000, 0001, 0010, 0011, 0100, 0101, 0110, 0111, 1000, 1010, 1100, 1110$\}$. Then the network state for this graph state can be constructed by 
\bea
N_{23}^{(Cl4)} = \sum_{\Vec{x} \in S} \frac{1}{12} \ketbra{\Vec{x}}_G^{(2)} \otimes \ketbra{\Vec{x}}_G^{(3)}.
\eea
It holds that
\bea
&& \tr[\rho_1 \otimes N_{23}^{(Cl4)} P^{(12)} \otimes ( \frac{1}{2}\mathbbm{1} - \ketbra{0000}_G ) ] \nonumber
\\ && \propto \tr[\rho W_{Cl4}],
\eea
where
\bea
P^{(12 )} &=& |\phi^{+}\rangle_{A_1A_2}\langle \phi^+| \otimes |\phi^{+}\rangle_{B_1B_2}\langle \phi^+| \nonumber
\\ && \otimes |\phi^{+}\rangle_{C_1C_2}\langle \phi^+| \otimes |\phi^{+}\rangle_{D_1D_2}\langle \phi^+|. \nonumber
\eea
Then a four-qubit entangled state $\rho$ is detected by $W_{Cl4}$ when
\bea
{}_{G} \langle 0000| \Lambda [\rho]| 0000\rangle_G > \frac{1}{2},
\eea
where the map $\Lambda^{(1 \to 3)}$ is defined as,
\bea
\Lambda^{( 1\rightarrow 3)} [\rho]= \frac{ \tr_{12}[\rho_{1}\otimes N_{23} P^{(12)}] }{ \tr[\rho_1\otimes N_{23} P^{(12)}]  }. \nonumber
\eea
\\

A typical decomposable EW can be written in the following form \cite{PhysRevA.84.032310}:
\bea
W_G = \frac{1}{2} \sum_{\Vec{x} \in S} \ketbra{\Vec{x}}_G - \ketbra{\Vec{0}}_G,
\eea
where the set $S \subseteq \{0,1\}^n$ depends on $W_G$.
Then the network states $N_{23}^{G}$ for this entanglement witnesses are given by
\bea
N_{23}^{(G)} = \sum_{\Vec{x} \in S} \frac{1}{|S|} \ketbra{\Vec{x}}_G^{(2)} \otimes \ketbra{\Vec{x}}_G^{(3)}.
\eea
An $n$-qubit entangled state $\rho$ is detected by $W_G$ when
\bea
{}_{G} \langle \Vec{0}| \Lambda [\rho]| \Vec{0}\rangle_G &>& \frac{1}{2},
\eea
for
\bea
\Lambda^{( 1\rightarrow 3)} [\rho] &=& \frac{ \tr_{12}[\rho_{1}\otimes N_{23} P^{(12)}] }{ \tr[\rho_1\otimes N_{23} P^{(12)}]  }, \nonumber\\ 
P^{(12)} &=& \Otimes_{v \in V} |\phi^+\rangle_{v_1 v_2}\langle\phi^+|, \nonumber
\eea
where $V$ denotes the set of vertices of the graph $G$, see Fig.~\ref{fig:figm}.

The overlap ${}_{G} \langle \Vec{0}| \Lambda [\rho]| \Vec{0}\rangle_G$ can be estimated with a fixed measurement on individual qubits. Note that a graph state is generated as follows, $\ket{\Vec{0}}_G$ is obtained as $\ket{\Vec{0}}_G = (\Otimes_{e \in E} U_{CZ}) (\Otimes_{v \in V} H) \ket{0}^{\otimes n}$. Then, the estimation of a singlet fraction can be achieved by applying an interaction $(\Otimes_{v \in V} H) (\Otimes_{e \in E} U_{CZ})$ to a resulting state $\Lambda^{(1 \to 3)} [\rho]$ and then performing a measurement in the computational basis.  The probability of obtaining an outcome $00$ gives the overlap ${}_{G} \langle \Vec{0}| \Lambda [\rho]| \Vec{0}\rangle_G$.
It holds that ${}_{G} \langle \Vec{0}| \Lambda [\rho]| \Vec{0}\rangle_G > \frac{1}{2}$ if and only if $\tr[\rho W] < 0$, which certifies that $\rho$ is entangled.

\end{document}